\documentclass[aps,preprint,nofootinbib,superscriptaddress]{revtex4}%
\usepackage{amssymb}
\usepackage{amsmath}
\usepackage{epsfig}
\usepackage{amsfonts}
\usepackage{graphicx}
\usepackage{hyperref}%
\providecommand{\U}[1]{\protect\rule{.1in}{.1in}}
\begin{document}
\title{Sheaf lines of Yang-Mills Instanton Sheaves}
\author{Sheng-Hong Lai}
\email{xgcj944137@gmail.com}
\affiliation{Department of Electrophysics, National Chiao-Tung University, Hsinchu, Taiwan, R.O.C.}
\author{Jen-Chi Lee}
\email{jcclee@cc.nctu.edu.tw}
\affiliation{Department of Electrophysics, National Chiao-Tung University, Hsinchu, Taiwan, R.O.C.}
\author{I-Hsun Tsai}
\email{ihtsai@math.ntu.edu.tw}
\affiliation{Department of Mathematics, National Taiwan University, Taipei, Taiwan, R.O.C. }
\date{\today}

\begin{abstract}
We calculate a sheaf line in $CP^{3}$ which is the real line supporting sheaf
points on $CP^{3}$ of $SL(2,C)$ Yang-Mills instanton (or $SU(2)$
\textit{complex} Yang-Mills instanton) sheaves for some given ADHM data we
obtained previously. We found that this sheaf line is indeed a special jumping
line over $S^{4}$ spacetime. In addition, we calculate the singularity
structure of the connection $A$ and the field strength $F$ at the
corresponding singular point on $S^{4}$ of this sheaf line. We found that the
order of singularity at the singular point on $S^{4}$ associated with the
sheaf line in $CP^{3}$ is higher than those of other singular points
associated with normal jumping lines. We conjecture that this is a general
feature for sheaf lines among jumping lines.

\end{abstract}
\maketitle
\tableofcontents

%

%TCIMACRO{\TeXButton{equation number}{\setcounter{equation}{0}
%\renewcommand{\theequation}{\arabic{section}.\arabic{equation}}}}%
%BeginExpansion
\setcounter{equation}{0}
\renewcommand{\theequation}{\arabic{section}.\arabic{equation}}%
%EndExpansion

\section{Introduction}

The discovery of classical Yang-Mills (YM) instanton began in 1975
\cite{BPST,CFTW,JR,Witten}. In a few years, the complete instanton solutions
with $8k-3$ moduli parameters for each $k$-th homotopy class were solved by
ADHM \cite{ADHM} in 1978 using theory in algebraic geometry. By using the
monad construction combining with the Penrose-Ward transform
\cite{ward1,ward2} , ADHM constructed the ADHM instanton solutions by
establishing an one to one correspondence between anti-self-dual
$SU(2)$-connections on $S^{4}$ and global holomorphic vector bundles of rank
two on $CP^{3}$. The explicit closed forms of the complete $SU(2)$ instanton
solutions with $k\leq3$ were calculated by physicists in \cite{CSW,JR2}. There
have been tremendous applications of YM instanton in quantum field theory
\cite{U(1),the} and geometry \cite{5} for the last few decades. For
references, see some review works in \cite{review}.

In a series of recent papers \cite{Ann,Ann2,Ann3}, instead of quaternion
calculation for the $SU(2)$ YM instanton, the present authors developed the
biquaternion method with biconjugation operation \cite{Ham} to construct
$SL(2,C)$ \cite{WY} YM instanton (or $SU(2)$ \textit{complex} YM instanton)
solutions with $16k-6$ moduli parameters for each $k$-th homotopy class. These
new $SL(2,C)$ instanton solutions contain previous $SL(2,C)$ $(M,N)$ instanton
solutions as a subset constructed in 1984 \cite{Lee}. The number of parameters
constructed in \cite{Ann} is consistent with the conjecture made by Frenkel
and Jardim in \cite{math2} and was proved recently in \cite{math3} from the
mathematical point of view \cite{math4,math5,math6,math7}.

Moreover, for the first time, in addition to the holomorphic vector bundles on
$CP^{3}$ in the ADHM construction which have been well studied in the $SU(2)$
instantons, the authors in \cite{Ann2,Ann3} discovered and explicitly
constructed the so-called YM instanton sheaves on $CP^{3}$. In contrast to the
smooth vector bundle on $CP^{3}$ induced by $SU(2)$ instanton on $S^{4}$ in
the Penrose-Ward transform \cite{ward1,ward2}, the vector bundle structure
breaks down and the dimension of vector space attached on $CP^{3}$ may vary
from point to point for $SL(2,C)$ YM instanton sheaves. In a previous
publication \cite{Ann2}, the authors calculated explicitly examples of sheaf
points on $CP^{3}$ where the dimension of the attached vector space changes.

Since there is a fibration of $CP^{3}$ down to $S^{4}$ with fiber being
$CP^{1}$ line \cite{Atiyah}, one important follow-up issue related to these
sheaf points is to study how to identify the corresponding points on $S^{4}$
spacetime. A related issue is to calculate the \textit{sheaf line }(see
Eq.(\ref{ppp})), or the real line in $CP^{3}$ which connecting the sheaf point
on $CP^{3}$ and the corresponding singular point on $S^{4}$. We will introduce
the Pl\"{u}cker coordinate to describe these sheaf lines in $CP^{3}$ in this paper.

Moreover, one would like to calculate the singularity structures of the
connection $A$ and the field strength $F$ on these singular points of $S^{4}$.
We will show that the order of singularity at the singular points on $S^{4}$
associated with sheaf line in $CP^{3}$ is higher than those of other singular
points associated with normal jumping lines. We conjecture that this is a
general feature for sheaf lines among jumping lines \cite{new}.

One unexpected result we obtained in our search of the sheaf lines was the
great simplification of the calculation of $v$ in Eq.(\ref{ttt}) and the
corresponding connection $A$ and the field strength $F$ associated with the
sheaf ADHM data. This is to be compared to the CFTW \cite{CFTW} real
$2$-instanton solution \cite{1977} which is lengthy and quite complicated. In
fact, we will see that for this sheaf ADHM data the explicit form of $SL(2,C)$
YM $2$-instanton field strength without removable singularities can be exactly calculated!

One important motivation to study $SL(2,C)$ (or in general $SL(n,C)$)
self-dual YM (SDYM) equations is the central role they play in the field of
integrable systems. Indeed, it was pointed out by Ward in 1985 \cite{ward3}
that many integrable or solvable systems with lower space dimension can be
obtained from $4D$ SDYM equations by reduction \cite{integr}. For the
geometric points of view, see \cite{mason1,mason2}. For the algebraic and
analytic aspects of this application, see \cite{integr}.

Some well known examples of this reduction are the KdV equation, the nonlinear
Schrodinger equation and the sine-Gordon equation. These integrable systems
contain classical solutions which were known as nontopological solitons,
envelope solitons and topological solitons respectively. All these soliton
solutions found numerous applications in many physical systems. See the book
\cite{book} for further references.

This paper is organized as following. In section two, we briefly review the
construction of YM instanton sheaves calculated in \cite{Ann2}. In section
three, we introduce Pl\"{u}cker coordinate to calculate jumping lines and
sheaf lines of YM $2$-instanton sheaves calculated in \cite{Ann2}. A duality
symmetry among YM instanton sheaf solutions was pointed out with application
to the known sheaf solutions. In section four, we calculate the singularity
structure of connection and field strength on $S^{4}$ spacetime associated
with jumping lines and sheaf lines of YM instanton sheaf. An explicit
$SL(2,C)$ YM $2$-ins$\tan$ton field strength will be exactly calculated. The
calculable exact $2$-ins$\tan$ton field strength is believed to be related to
the $2$-instanton sheaf structure. A conclusion is presented in section five.

\section{The $SL(2,C)$ Yang-Mills two instanton sheaves}

In this section, we briefly review the biquaternion construction of $SL(2,C)$
ADHM instantons \cite{Ann,Ann2}. We will pay attention to the existence of
jumping lines and sheaf structures of YM $2$-instanton sheaves
\cite{Ann,Ann2,Ann3}.

\subsection{The biquaternion construction of $SL(2,C)$ ADHM instantons}

In the biquaternion construction of $SL(2,C)$ ADHM instanton, the quadratic
condition on the biquaternion matrix $\Delta(x)$ of $SL(2,C)$ instantons reads
\cite{CSW,Ann,Ann2}%

\begin{equation}
\Delta(x)^{\circledast}\Delta(x)=f^{-1}=\text{symmetric, non-singular }k\times
k\text{ matrix for }x\notin J \label{ff}%
\end{equation}
where for $x\in J,$%
\begin{equation}
\det\Delta(x)^{\circledast}\Delta(x)=0. \label{points}%
\end{equation}
The set $J$ is called singular locus or "jumping lines". There are no singular
locus for $SU(2)$ instantons on $S^{4}$ \cite{CSW}. The biconjugation
\cite{Ham} of a biquaternion%
\begin{equation}
z=z_{\mu}e_{\mu}\text{, \ }z_{\mu}\in C,
\end{equation}
is defined to be%
\begin{equation}
z^{\circledast}=z_{\mu}e_{\mu}^{\dagger}=z_{0}e_{0}-z_{1}e_{1}-z_{2}%
e_{2}-z_{3}e_{3}=x^{\dagger}+y^{\dagger}i. \label{x}%
\end{equation}
Occasionally the unit quarternions can be expressed as Pauli matrices%
\begin{equation}
e_{0}\rightarrow%
\begin{pmatrix}
1 & 0\\
0 & 1
\end{pmatrix}
,e_{i}\rightarrow-i\sigma_{i}\ \text{; }i=1,2,3.
\end{equation}
The norm square of a biquarternion is defined to be%
\begin{equation}
|z|_{c}^{2}=z^{\circledast}z=(z_{0})^{2}+(z_{1})^{2}+(z_{2})^{2}+(z_{3})^{2},
\end{equation}
which is a \textit{complex} number in general.

As a simple example, for the case of $SL(2,C)$ diagonal CFTW $2$-instanton%
\begin{equation}
\Delta(x)=%
\begin{bmatrix}
\lambda_{1} & \lambda_{2}\\
x-y_{1} & 0\\
0 & x-y_{2}%
\end{bmatrix}
,
\end{equation}

\begin{equation}
\Delta^{\circledast}(x)=%
\begin{bmatrix}
\lambda_{1}^{\circledast} & x^{\circledast}-y_{1}^{\circledast} & 0\\
\lambda_{2}^{\circledast} & 0 & x^{\circledast}-y_{2}^{\circledast}%
\end{bmatrix}
\end{equation}
where in the ADHM data $\lambda_{j}$ a \textit{complex} number, and $y_{j}$ a biquaternion.

One can calculate the gauge potential as \cite{Ann}%
\begin{align}
A_{\mu}  &  =v^{\circledast}\partial_{\mu}v=\frac{1}{4}[e_{\mu}^{\dagger
}e_{\nu}-e_{\nu}^{\dagger}e_{\mu}]\partial_{\nu}\ln(1+\frac{\lambda_{1}^{2}%
}{|x-y_{1}|_{c}^{2}}+\frac{\lambda_{2}^{2}}{|x-y_{k}|_{c}^{2}})\nonumber\\
&  =\frac{1}{4}[e_{\mu}^{\dagger}e_{\nu}-e_{\nu}^{\dagger}e_{\mu}%
]\partial_{\nu}\ln(\phi)
\end{align}
where%

\begin{equation}
v=\frac{1}{\sqrt{\phi}}%
\begin{bmatrix}
1\\
-\frac{\lambda_{1}(x_{\mu}-y_{1\mu})e_{\mu}}{|x-y_{1}|^{2}}\\
-\frac{\lambda_{2}(x_{\mu}-y_{2\mu})e_{\mu}}{|x-y_{2}|^{2}}%
\end{bmatrix}
\end{equation}
and%
\begin{equation}
\phi=1+\frac{\lambda_{1}^{2}}{|x-y_{1}|_{c}^{2}}+\frac{\lambda_{2}^{2}%
}{|x-y_{2}|_{c}^{2}}.
\end{equation}
To get the non-removable singularities or jumping lines, it turned out that
one needs to calculate zeros of the normalization factor $\phi$ \cite{Ann}%
\begin{align}
|x-y_{1}|_{c}^{2}|x-y_{2}|_{c}^{2}\phi &  =|x-y_{1}|_{c}^{2}|x-y_{2}|_{c}%
^{2}+|\lambda_{2}|_{c}^{2}|x-y_{1}|_{c}^{2}+|\lambda_{1}|_{c}^{2}|x-y_{2}%
|_{c}^{2}\nonumber\\
&  =P_{4}(x)+iP_{3}(x)=0. \label{zero}%
\end{align}
For the $SL(2,C)$ CFTW general $k$-instanton case, one encounters
intersections of zeros of $P_{2k}(x)$ and $P_{2k-1}(x)$ polynomials with
degrees $2k$ and $2k-1$ respectively%
\begin{equation}
P_{2k}(x)=0,\text{ \ }P_{2k-1}(x)=0.
\end{equation}
One notes that Eq.(\ref{zero}) can be written as%
\begin{equation}
\det\Delta(x)^{\circledast}\Delta(x)=|x-y_{1}|_{c}^{2}|x-y_{2}|_{c}%
^{2}+|\lambda_{2}|_{c}^{2}|x-y_{1}|_{c}^{2}+|\lambda_{1}|_{c}^{2}|x-y_{2}%
|_{c}^{2}=0
\end{equation}
which gives the jumping lines of the $SL(2,C)$ diagonal CFTW $2$-instanton. It
was shown that \cite{Ann2} there is no sheaf line structure for the $SL(2,C)$
diagonal CFTW $k$-instanton. The complete jumping lines of ADHM $2$-instanton
and $3$-instanton of Eq.(\ref{points}) were calculated in \cite{Ann}. However,
the existence of sheaf lines was not known and not calculated there. We will
calculate and identify some sheaf lines of the $SL(2,C)$ ADHM $2$-instanton in
the next section.

\subsection{The $SL(2,C)$ complex ADHM equations}

The second method to construct $SL(2,C)$ ADHM data is to solve the complex
ADHM equations \cite{Donald}%
\begin{subequations}
\begin{align}
\left[  B_{11},B_{12}\right]  +I_{1}J_{1}  &  =0,\label{adhm1}\\
\left[  B_{21},B_{22}\right]  +I_{2}J_{2}  &  =0,\label{adhm2}\\
\left[  B_{11},B_{22}\right]  +\left[  B_{21},B_{12}\right]  +I_{1}J_{2}%
+I_{2}J_{1}  &  =0. \label{adhm3}%
\end{align}
In this approach, one key step is to use the explicit matrix representation
(EMR) \cite{Ann2} of the biquaternion and do the rearrangement rule
\cite{Ann2} to \bigskip explicitly identify the complex ADHM data
$(B_{lm},I_{m,}J_{m})$ with $l,m=1,2$ from the $\Delta(x)$ matrix in
Eq.(\ref{ff}).

As an explicit example and for illustration, we calculate the $SL(2,C)$ CFTW
$2$-instanton case. In the EMR, a biquaternion can be written as a $2\times2$
complex matrix
\end{subequations}
\begin{align}
z  &  =z^{0}e_{0}+z^{1}e_{1}+z^{2}e_{2}+z^{3}e_{3}\nonumber\\
&  =%
\begin{bmatrix}
\left(  a^{0}+b^{3}\right)  +i\left(  b^{0}-a^{3}\right)  & \left(
-a^{2}+b^{1}\right)  +i\left(  -b^{2}-a^{1}\right) \\
\left(  a^{2}+b^{1}\right)  +i\left(  b^{2}-a^{1}\right)  & \left(
a^{0}-b^{3}\right)  +i\left(  b^{0}+a^{3}\right)
\end{bmatrix}
\end{align}
where $a^{\mu}$ and $b^{\mu}$ are real and imaginary parts of $z^{\mu}$
respectively. For the CFTW $2$-instanton case%

\begin{align}
a  &  =%
\begin{bmatrix}
\lambda_{1} & \lambda_{2}\\
y_{11} & 0\\
0 & y_{22}%
\end{bmatrix}
=%
\begin{bmatrix}
p_{1}+iq_{1} & 0 & p_{2}+iq_{2} & 0\\
0 & p_{1}+iq_{1} & 0 & p_{2}+iq_{2}\\
y_{11}^{0}-iy_{11}^{3} & -\left(  y_{11}^{2}+iy_{11}^{1}\right)  & 0 & 0\\
y_{11}^{2}-iy_{11}^{1} & y_{11}^{0}+iy_{11}^{3} & 0 & 0\\
0 & 0 & y_{22}^{0}-iy_{22}^{3} & -\left(  y_{22}^{2}+iy_{22}^{1}\right) \\
0 & 0 & y_{22}^{2}-iy_{22}^{1} & y_{22}^{0}+iy_{22}^{3}%
\end{bmatrix}
\\
&  \rightarrow%
\begin{bmatrix}
p_{1}+iq_{1} & p_{2}+iq_{2} & 0 & 0\\
0 & 0 & p_{1}+iq_{1} & p_{2}+iq_{2}\\
y_{11}^{0}-iy_{11}^{3} & 0 & -\left(  y_{11}^{2}+iy_{11}^{1}\right)  & 0\\
0 & y_{22}^{0}-iy_{22}^{3} & 0 & -\left(  y_{22}^{2}+iy_{22}^{1}\right) \\
y_{11}^{2}-iy_{11}^{1} & 0 & y_{11}^{0}+iy_{11}^{3} & 0\\
0 & y_{22}^{2}-iy_{22}^{1} & 0 & y_{22}^{0}+iy_{22}^{3}%
\end{bmatrix}
=%
\begin{bmatrix}
J_{1} & J_{2}\\
B_{11} & B_{21}\\
B_{12} & B_{22}%
\end{bmatrix}
\label{2nd}%
\end{align}
where in Eq.(\ref{2nd}) we have done the \textit{rearrangement rule }for an
element $z_{ij}$ in $a$\textit{ }
\begin{align}
z_{2n-1,2m-1}  &  \rightarrow z_{n,m}\text{ },\nonumber\\
z_{2n-1,2m}  &  \rightarrow z_{n,k+m}\text{ },\nonumber\\
z_{2n,2m-1}  &  \rightarrow z_{k+n,m}\text{ },\nonumber\\
z_{2n,2m}  &  \rightarrow z_{k+n,k+m}.
\end{align}
The EMR and the rearrangement rule for $a^{\circledast}$ can be similarly performed.

For the $SU(2)$ ADHM instantons, one imposes the conditions%
\begin{subequations}
\begin{align}
I_{1}  &  =J^{\dagger},I_{2}=-I,J_{1}=I^{\dagger},J_{2}=J,\nonumber\\
B_{11}  &  =B_{2}^{\dagger},B_{12}=B_{1}^{\dagger},B_{21}=-B_{1},B_{22}=B_{2}%
\end{align}
to recover the real ADHM equations%
\end{subequations}
\begin{subequations}
\begin{align}
\left[  B_{1},B_{2}\right]  +IJ  &  =0,\\
\left[  B_{1},B_{1}^{\dagger}\right]  +\left[  B_{2},B_{2}^{\dagger}\right]
+II^{\dagger}-J^{\dagger}J  &  =0.
\end{align}

\subsection{The monad construction and YM $2$-instanton sheaves}

The third method to construct $SL(2,C)$ ADHM instanton is the monad
construction \cite{ADHM}. This method is particular suitable for constructing
instanton sheaves. One introduces the $\bigskip\alpha\,$and $\beta$ matrices
as functions of homogeneous coordinates $z,w,x,y$ of $CP^{3}$ and defines%
\end{subequations}
\begin{subequations}
\begin{align}
\alpha &  =%
\begin{bmatrix}
zB_{11}+wB_{21}+x\\
zB_{12}+wB_{22}+y\\
zJ_{1}+wJ_{2}%
\end{bmatrix}
,\label{alpha}\\
\beta &  =%
\begin{bmatrix}
-zB_{12}-wB_{22}-y & zB_{11}+wB_{21}+x & zI_{1}+wI_{2}%
\end{bmatrix}
. \label{beta}%
\end{align}
It can be shown that \cite{Ann2}\ the condition%
\end{subequations}
\begin{equation}
\beta\alpha=0
\end{equation}
is satisfied if and only if the complex ADHM equations in Eq.(\ref{adhm1}) to
Eq.(\ref{adhm3}) holds.

In the monad construction of the holomorphic vector bundles, either $\beta$ is
not surjective (or onto) or $\alpha$ is not injective (or 1-1) at some points
of $CP^{3}$ for some ADHM data, the dimension of $($Ker $\beta$/ Im $\alpha)$
varies from point to point on $CP^{3}$, and one encounters "instanton sheaves"
on $CP^{3}$ \cite{math2}. In our previous publication \cite{Ann2}, we
discovered that for some ADHM data at some sheaf points on $CP^{3}$, there
exists common eigenvector $u$ in the costable condition $\alpha u=0$ or
\cite{math2}
\begin{subequations}
\begin{align}
\left(  zB_{11}+wB_{21}\right)  u  &  =-xu,\label{a}\\
\left(  zB_{12}+wB_{22}\right)  u  &  =-yu,\label{b}\\
\left(  zJ_{1}+wJ_{2}\right)  u  &  =0. \label{c}%
\end{align}
So $\alpha$ is not injective there and the dimension of $($Ker $\beta$/ Im
$\alpha)$ is not a constant over $CP^{3}$.

The first example of YM instanton sheaf discovered in \cite{Ann2} was the
$2$-instanton sheaf. For points $[x:y:z:w]=[0:0:1:\pm1]$ on $CP^{3}$ with the
ADHM data\bigskip%
\end{subequations}
\begin{equation}%
\begin{bmatrix}
\lambda_{1} & \lambda_{2}\\
y_{11} & y_{12}\\
y_{12} & y_{22}%
\end{bmatrix}
=%
\begin{bmatrix}
a & 0 & 0 & ia\\
0 & a & ia & 0\\
\frac{-i}{\sqrt{2}}a & 0 & 0 & \frac{a}{\sqrt{2}}\\
0 & \frac{-i}{\sqrt{2}}a & \frac{a}{\sqrt{2}} & 0\\
0 & \frac{a}{\sqrt{2}} & \frac{i}{\sqrt{2}}a & 0\\
\frac{a}{\sqrt{2}} & 0 & 0 & \frac{i}{\sqrt{2}}a
\end{bmatrix}
,a\in C,a\neq0, \label{data1}%
\end{equation}
$\alpha$ is not injective. The second example of YM $2$-instanton sheaf
discovered \cite{Ann2} was for points $[x:y:z:w]=[0:0:1:\pm i]$ on $CP^{3}$
with the ADHM data%
\begin{equation}%
\begin{bmatrix}
\lambda_{1} & \lambda_{2}\\
y_{11} & y_{12}\\
y_{12} & y_{22}%
\end{bmatrix}
=%
\begin{bmatrix}
a & 0 & 0 & a\\
0 & a & -a & 0\\
\frac{-i}{\sqrt{2}}a & 0 & 0 & \frac{-ia}{\sqrt{2}}\\
0 & \frac{-i}{\sqrt{2}}a & \frac{ia}{\sqrt{2}} & 0\\
0 & \frac{-ia}{\sqrt{2}} & \frac{i}{\sqrt{2}}a & 0\\
\frac{ia}{\sqrt{2}} & 0 & 0 & \frac{i}{\sqrt{2}}a
\end{bmatrix}
,a\in C,a\neq0, \label{data2}%
\end{equation}
$\alpha$ is not injective.

\section{Jumping lines and sheaf lines of instanton sheaves}

In the previous section, we have obtained sheaf points on $CP^{3}$ with some
examples of given ADHM data. One natural issue arose then is to study how to
identify the corresponding points on $S^{4}$ and calculate the singularity
structure of the connection $A$ and the field strength $F$ on these points.
The latter issue will be studied in the next section. In this section, we
first define and calculate the \textit{sheaf line}, or the real line which
connecting the sheaf point on $CP^{3}$ and the corresponding singular point on
$S^{4}$ and see whether the sheaf line\ is a jumping line or not. (see
Eq.(\ref{ppp} in section III.D)

In our previous publication \cite{Ann2}, we have shown that there is no sheaf
line structure for the $SL(2,C)$ diagonal CFTW $k$-instanton. On the other
hand, the complete jumping lines of ADHM $2$-instanton and $3$-instanton of
Eq.(\ref{points}) were calculated in section IV D. of \cite{Ann}. However, the
existence of sheaf lines was not known and not calculated there. In this
section, we will calculate and identify some sheaf line of the $SL(2,C)$ ADHM
$2$-instanton.

\subsection{Real lines in $CP^{3}$}

It is well known that there is a fibration from $CP^{3}$ to $S^{4}$ with
fibers being $CP^{1}$ \cite{Atiyah}. In the Pl\"{u}cker coordinate
representation of a (complex) line $CP^{1}$ in $CP^{3}$, one uses six
homogeneous coordinates to represent each line. More specifically, given two
points $[a:b:c:d]$ and $[x:y:z:w]$ on $CP^{3}$, the Pl\"{u}cker coordinates
$z_{ij}$ of the line $L$ connecting the two points are defined
as\cite{ward2,Atiyah}
\begin{align}
z_{12}  &  =ay-bx,\nonumber\\
z_{13}  &  =az-cx,\nonumber\\
z_{14}  &  =aw-dx,\nonumber\\
z_{23}  &  =bz-cy,\nonumber\\
z_{24}  &  =bw-dy,\nonumber\\
z_{34}  &  =cw-dz, \label{plucker}%
\end{align}
or in short%
\begin{equation}
\lbrack z_{12}:z_{13}:z_{14}:z_{23}:z_{24}:z_{34}]=[a:b:c:d]\wedge\lbrack
x:y:z:w].
\end{equation}

Note that the Pl\"{u}cker coordinates defined above are uniquely determined by
$L$ up to a common nonzero factor and not all six components can be zero. Thus
$z_{ij}$ can be thought of as homogeneous coordinates of a point in $CP^{5}$.
However, not all points in $CP^{5}$ correspond to lines in $CP^{3}.$ The
Pl\"{u}cker coordinates of a line satisfy the quadratic relations
\cite{Griffiths}%
\begin{equation}
z_{12}z_{34}+z_{13}z_{42}+z_{14}z_{23}=0, \label{rel}%
\end{equation}
as can be easily verified from the definition in Eq.(\ref{plucker}). So the
set of lines in $CP^{3}$ constitutes a manifold of complex dimension $4$
rather than $5$.

A line in $CP^{3}$ is called a \textit{real line} if it is a fiber on $S^{4}$.
To characterize a real line in $CP^{3}$, one introduces the $\sigma$ map which
preserves a real line $L$ \cite{Atiyah}
\begin{equation}
\sigma(L)=L\text{ if and only if }L=\text{ }real\text{ }line\text{.}
\label{sigma}%
\end{equation}
The $\sigma$ map can be defined as following. Let $\pi$ be the projection of
the fibration from $CP^{3}$ to $S^{4}$\bigskip\ \cite{Atiyah}%
\begin{equation}
\pi\text{ }:CP^{3}\text{ }\rightarrow S^{4}\cong\text{ }HP^{1} \label{hp}%
\end{equation}
where $HP^{1}$ is the quaternion projective space. We can parametrize the
projection $\pi$ as \cite{Atiyah}%
\begin{equation}
\pi\text{ }:[z_{1}:z_{2}:z_{3}:z_{4}]\rightarrow\lbrack z_{1}+z_{2}%
j:z_{3}+z_{4}j]
\end{equation}
where $j\equiv e_{2}$ is a unit quaternion defined in Eq.(\ref{x}). The
$\sigma$ map in Eq.(\ref{sigma}) can then be written as \cite{Atiyah}%
\begin{equation}
\sigma:[z_{1}:z_{2}:z_{3}:z_{4}]\rightarrow\lbrack\overline{z}_{2}%
:-\overline{z}_{1}:\overline{z}_{4}:-\overline{z}_{3}].
\end{equation}

It can be shown that \cite{Atiyah} the $\sigma$ map preserves real lines as
illustrated in Eq.(\ref{sigma}) or%
\begin{equation}
\pi\circ\sigma=\pi\text{ }. \label{map}%
\end{equation}
In fact (we use the notation $(1,i,j,k)=(e_{0},e_{1},e_{2},e_{3})$)%
\begin{align}
\pi([x  &  :y:z:w])=[x+ye_{2}:z+we_{2}]\nonumber\\
&  =[x^{0}e_{0}+x^{1}e_{1}+(y^{0}e_{0}+y^{1}e_{1})e_{2}:z^{0}e_{0}+z^{1}%
e_{1}+(w^{0}e_{0}+w^{1}e_{1})e_{2}]\nonumber\\
&  =[x^{0}e_{0}+x^{1}e_{1}+y^{0}e_{2}+y^{1}e_{3}:z^{0}e_{0}+z^{1}e_{1}%
+w^{0}e_{2}+w^{1}e_{3}]. \label{ii}%
\end{align}
In Eq.(\ref{ii}), $x^{0}$ and $x^{1}$ are the real part and imaginary part of
the complex number $x=x^{0}e_{0}+x^{1}e_{1}=x^{0}+x^{1}\sqrt{-1},$ $etc$. On
the other hand%
\begin{align}
\pi\circ\sigma\lbrack x  &  :y:z:w]=\pi([\bar{y}:-\bar{x}:\bar{w}:-\bar
{z}])\nonumber\\
&  =\pi([y^{0}e_{0}-y^{1}e_{1}:-x^{0}e_{0}+x^{1}e_{1}:w^{0}e_{0}-w^{1}%
e_{1}:-z^{0}e_{0}+z^{1}e_{1}])\nonumber\\
&  =[y^{0}e_{0}-y^{1}e_{1}+(-x^{0}e_{0}+x^{1}e_{1})e_{2}:w^{0}e_{0}-w^{1}%
e_{1}+(-z^{0}e_{0}+z^{1}e_{1})e_{2}]\nonumber\\
&  =[y^{0}e_{0}-y^{1}e_{1}-x^{0}e_{2}+x^{1}e_{3}:w^{0}e_{0}-w^{1}e_{1}%
-z^{0}e_{2}+z^{1}e_{3}]\nonumber\\
&  \simeq e_{2}[y^{0}e_{0}-y^{1}e_{1}-x^{0}e_{2}+x^{1}e_{3}:w^{0}e_{0}%
-w^{1}e_{1}-z^{0}e_{2}+z^{1}e_{3}]\nonumber\\
&  =[x^{0}e_{0}+x^{1}e_{1}+y^{0}e_{2}+y^{1}e_{3}:z^{0}e_{0}+z^{1}e_{1}%
+w^{0}e_{2}+w^{1}e_{3}]\nonumber\\
&  =\pi([x:y:z:w]),
\end{align}
which proves Eq.(\ref{map}).

\subsection{A duality symmetry}

With the $\sigma$ map introduced in the previous subsection, we can show an
important duality symmetry \cite{misleading} among instanton sheaf solutions.
In \cite{math2}, it was noted that given a set of ADHM data, one can generate
a new set of ADHM data through the map%
\begin{equation}
\Sigma:(B_{11},B_{12},B_{21},B_{22},I_{1},I_{2},J_{1},J_{2})\rightarrow
(B_{22}^{+},-B_{21}^{+},-B_{12}^{+},B_{11}^{+},J_{2}^{+},-J_{1}^{+},-I_{2}%
^{+},I_{1}^{+}).
\end{equation}
Recall that in the monad construction of instanton bundle, if either $\alpha$
is not injective or $\beta$ is not surjective, then the dimension of $($Ker
$\beta$/ Im $\alpha)$ may vary from point to point on $CP^{3}$, and one is led
to use sheaf description for YM instantons or "instanton sheaves" on $CP^{3}$.
The costable conditions in Eq.(\ref{a}) to Eq.(\ref{c}) or%
\begin{equation}
\alpha u=0.
\end{equation}
imply $\alpha$ is not injective. another choice is%
\begin{equation}
\beta^{+}u=0
\end{equation}
or the stable condition \cite{math2}%
\begin{subequations}
\begin{align}
\left(  \overline{z}B_{11}^{+}+\overline{w}B_{21}^{+}\right)  u  &
=-\overline{x}u,\label{d}\\
\left(  \overline{z}B_{12}^{+}+\overline{w}B_{22}^{+}\right)  u  &
=-\overline{y}u,\label{e}\\
\left(  \overline{z}I_{1}^{+}+\overline{w}I_{2}^{+}\right)  u  &  =0 \label{f}%
\end{align}
which imply $\beta$ is not surjective. One notes that by applying the $\Sigma$
map on the ADHM data and the $\sigma$ map on the point $(x,y,z,w)$ on $CP^{3}%
$, Eq.(\ref{a}) to Eq.(\ref{c}) are transformed to Eq.(\ref{d}) to Eq.(\ref{f}).

To be more precisely, with a sheaf solution of Eq.(\ref{a}) to Eq.(\ref{c}),
one can define a set of new ADHM data%
\end{subequations}
\begin{equation}
(B_{11}^{\prime},B_{12}^{\prime},B_{21}^{\prime},B_{22}^{\prime},I_{1}%
^{\prime},I_{2}^{\prime},J_{1}^{\prime},J_{2}^{\prime})=(B_{22}^{+}%
,-B_{21}^{+},-B_{12}^{+},B_{11}^{+},J_{2}^{+},-J_{1}^{+},-I_{2}^{+},I_{1}%
^{+}), \label{sol1}%
\end{equation}
and at the new point%
\begin{equation}
\lbrack x^{\prime},y^{\prime},z^{\prime},w^{\prime}]=[\overline{y}%
:-\overline{x}:\overline{w}:-\overline{z}] \label{sol2}%
\end{equation}
on $CP^{3}$. One can verify that Eq.(\ref{sol1}) together with Eq.(\ref{sol2})
constitute a new sheaf solution of Eq.(\ref{d}) to Eq.(\ref{f}). In fact, if
one plugs Eq.(\ref{sol1}) and Eq.(\ref{sol2}) into Eq.(\ref{d}) to
Eq.(\ref{f}), one ends up with precisely Eq.(\ref{a}) to Eq.(\ref{c}). That
is, $\alpha$ is not injective for the old sheaf solution and $\beta^{\prime}$
is not surjective for the new sheaf solution.

As an example of the dual symmetry discussed above, we use the old sheaf point
$\left[  x:y:z:w\right]  =\left[  0:0:1:1\right]  $ with the old ADHM data in
Eq.(\ref{data1})
\begin{align}
B_{11}  &  =%
\begin{pmatrix}
\frac{-ia}{\sqrt{2}} & 0\\
0 & \frac{ia}{\sqrt{2}}%
\end{pmatrix}
,B_{21}=%
\begin{pmatrix}
0 & \frac{a}{\sqrt{2}}\\
\frac{a}{\sqrt{2}} & 0
\end{pmatrix}
,B_{12}=%
\begin{pmatrix}
0 & \frac{a}{\sqrt{2}}\\
\frac{a}{\sqrt{2}} & 0
\end{pmatrix}
,B_{22}=%
\begin{pmatrix}
\frac{-ia}{\sqrt{2}} & 0\\
0 & \frac{ia}{\sqrt{2}}%
\end{pmatrix}
,\nonumber\\
J_{1}  &  =%
\begin{pmatrix}
a & 0\\
0 & ia
\end{pmatrix}
,J_{2}=%
\begin{pmatrix}
0 & ia\\
a & 0
\end{pmatrix}
,I_{1}=%
\begin{pmatrix}
0 & a\\
-ia & 0
\end{pmatrix}
,I_{2}=%
\begin{pmatrix}
-a & 0\\
0 & ia
\end{pmatrix}
,
\end{align}
which give $\alpha$ not injective, then we can calculate the new sheaf point
$\left[  x^{\prime}:y^{\prime}:z^{\prime}:w^{\prime}\right]  =[\bar{y}%
:-\bar{x}:\bar{w}:-\bar{z}]=\left[  0:0:1:-1\right]  $ with the new ADHM data
\begin{align}
B_{11}^{\prime}  &  =%
\begin{pmatrix}
\frac{i\bar{a}}{\sqrt{2}} & 0\\
0 & \frac{-i\bar{a}}{\sqrt{2}}%
\end{pmatrix}
,B_{21}^{\prime}=%
\begin{pmatrix}
0 & -\frac{\bar{a}}{\sqrt{2}}\\
-\frac{\bar{a}}{\sqrt{2}} & 0
\end{pmatrix}
,B_{12}^{\prime}=%
\begin{pmatrix}
0 & -\frac{\bar{a}}{\sqrt{2}}\\
-\frac{\bar{a}}{\sqrt{2}} & 0
\end{pmatrix}
,B_{22}^{\prime}=%
\begin{pmatrix}
\frac{i\bar{a}}{\sqrt{2}} & 0\\
0 & \frac{-i\bar{a}}{\sqrt{2}}%
\end{pmatrix}
,\nonumber\\
J_{1}^{\prime}  &  =%
\begin{pmatrix}
\bar{a} & 0\\
0 & i\bar{a}%
\end{pmatrix}
,J_{2}^{\prime}=%
\begin{pmatrix}
0 & i\bar{a}\\
\bar{a} & 0
\end{pmatrix}
,I_{1}^{\prime}=%
\begin{pmatrix}
0 & \bar{a}\\
-i\bar{a} & 0
\end{pmatrix}
,I_{2}^{\prime}=%
\begin{pmatrix}
-\bar{a} & 0\\
0 & i\bar{a}%
\end{pmatrix}
, \label{data3}%
\end{align}
which give $\beta^{\prime}$ not surjective. Note that at the point $\left[
0:0:1:-1\right]  $ with the ADHM data in Eq.(\ref{data1}), $\alpha$ is not
injective. It's important to see that the ADHM data in Eq.(\ref{data3}) can
not be obtained from the ADHM data in Eq.(\ref{data1}) by re-naming the
parameter $a$.

\subsection{Jumping lines}

In contrast to the $SU(2)$ ADHM construction, the $SL(2,C)$ ADHM instanton
construction in Eq.(\ref{ff}) contains\ a set of jumping lines $J$ for the
instanton bundle $E$. At those spacetime points $x\in J\subset S^{4}$ with
$\det\Delta(x)^{\circledast}\Delta(x)=0$, the connections $A$ and the field
strength $F$ are singular (see Eq.(\ref{f2}) and Eq.(\ref{f8}) in section IV.C).

On the other hand, the real lines which connect points $[a:b:c:d]$ and
$[x:y:z:w]$ on $CP^{3}$ are jumping lines of the instanton bundle $E$ if
$\det(\beta_{\lbrack a:b:c:d]}\alpha_{\lbrack x:y:z:w]})=0$ (see
Eq.(\ref{id5}) and Eq.(\ref{id1}) in this section). It turns out that there is
an one to one correspondence between jumping lines of the instanton bundle $E$
and singular points of $A$ and $F$ on $S^{4}$ spacetime (see Eq.(\ref{id6} in
this section).

Note that a bundle $E$ on $CP^{3}$ can descend down to a bundle over $S^{4}$
if and only if no fiber of the twistor fibration is a jumping line for $E$.
This is the case for $SU(2)$ instanton and thus there are no jumping lines on
$E$ and no singular points on $S^{4}$ spacetime.

\bigskip To see the correspondence, similar to Eq.(\ref{alpha}) and
Eq.(\ref{beta}), we define $\alpha$ and $\beta$ matrices at different points
$[x:y:z:w]$ and $[a:b:c:d]$ on $CP^{3}$ as%

\begin{align}
\alpha_{\lbrack x:y:z:w]}  &  =%
\begin{pmatrix}
I_{2\times2}\\
0_{2\times2}\\
0_{2\times2}%
\end{pmatrix}
x+%
\begin{pmatrix}
0_{2\times2}\\
I_{2\times2}\\
0_{2\times2}%
\end{pmatrix}
y+%
\begin{pmatrix}
B_{11}\\
B_{12}\\
J_{1}%
\end{pmatrix}
z+%
\begin{pmatrix}
B_{21}\\
B_{22}\\
J_{2}%
\end{pmatrix}
w,\\
\beta_{\lbrack a:b:c:d]}  &  =%
\begin{pmatrix}
0_{2\times2} & I_{2\times2} & 0_{2\times2}%
\end{pmatrix}
a+%
\begin{pmatrix}
-I_{2\times2} & 0_{2\times2} & 0_{2\times2}%
\end{pmatrix}
b\nonumber\\
&  +%
\begin{pmatrix}
-B_{12} & B_{11} & I_{1}%
\end{pmatrix}
c+%
\begin{pmatrix}
-B_{22} & B_{21} & I_{2}%
\end{pmatrix}
d.
\end{align}
It is straightforward to calculate the product map
\begin{align}
&  \beta_{\lbrack a:b:c:d]}\alpha_{\lbrack x:y:z:w]}\nonumber\\
&  =\left(  ay-bx\right)  +B_{12}\left(  az-cx\right)  +B_{22}\left(
aw-dx\right)  +B_{11}\left(  cy-bz\right)  +B_{21}\left(  dy-bw\right)
\nonumber\\
&  +\left(  -B_{12}B_{11}+B_{11}B_{12}+i_{1}j_{1}\right)  cz+\left(
-B_{12}B_{21}+B_{11}B_{22}+i_{1}j_{2}\right)  cw\nonumber\\
&  +\left(  -B_{22}B_{11}+B_{21}B_{12}+i_{2}j_{1}\right)  dz+\left(
-B_{22}B_{21}+B_{21}B_{22}+i_{2}j_{2}\right)  dw\nonumber\\
&  =z_{12}+B_{12}z_{13}+B_{22}z_{14}-B_{11}z_{23}-B_{21}z_{24}+\left(
-B_{12}B_{21}+B_{11}B_{22}+i_{1}j_{2}\right)  z_{34}%
\end{align}
where we have applied the complex ADHM equations in Eq.(\ref{adhm1}) to
Eq.(\ref{adhm3}). We have also used the Pl\"{u}cker coordinate representation
in Eq.(\ref{plucker}) to reduce the above result.

As an example, for the sheaf point $[0:0:1:1]$ on $CP^{3}$ obtained in the
previous section, we can calculate%
\begin{equation}
\beta_{\lbrack0:0:1:1]}\alpha_{\sigma\lbrack0:0:1:1]}=\beta_{\lbrack
0:0:1:1]}\alpha_{\lbrack0:0:1:-1]}=%
\begin{pmatrix}
0 & 0\\
0 & 0
\end{pmatrix}
.
\end{equation}

On the other hand, we can also calculate $\Delta^{\circledast}\Delta$ in
Eq.(\ref{ff}) on $S^{4}$. To do the calculation, we introduce the coordinates
for $x$ ($x_{0}$ and $x_{1}$ in Eq.(\ref{point}) are not to be confused with
$x^{0}$ and $x^{1}$ in Eq.(\ref{ii}))%
\begin{align}
x  &  =x_{0}e_{0}+x_{1}e_{1}+x_{2}e_{2}+x_{3}e_{3}\nonumber\\
&  =%
\begin{pmatrix}
x_{0}-ix_{3} & -\left(  x_{2}+ix_{1}\right) \\
x_{2}-ix_{1} & x_{0}+ix_{3}%
\end{pmatrix}
\nonumber\\
&  =%
\begin{pmatrix}
x_{11} & x_{21}\\
x_{12} & x_{22}%
\end{pmatrix}
. \label{point}%
\end{align}
The result is%
\begin{align}
&  \Delta^{\circledast}\Delta\nonumber\\
&  =%
\begin{pmatrix}
-I_{2} & x_{22}+B_{22} & -x_{21}-B_{21}\\
I_{1} & -x_{12}-B_{12} & x_{11}+B_{11}%
\end{pmatrix}%
\begin{pmatrix}
J_{1} & J_{2}\\
x_{11}+B_{11} & x_{21}+B_{21}\\
x_{12}+B_{12} & x_{22}+B_{22}%
\end{pmatrix}
\nonumber\\
&  =\left(  x_{11}x_{22}-x_{12}x_{21}+x_{11}B_{22}-x_{12}B_{21}-x_{21}%
B_{12}+x_{22}B_{11}+I_{1}J_{2}+B_{11}B_{22}-B_{12}B_{21}\right)
\end{align}
where again the complex ADHM equations have been used to reduce the
calculation above.

Finally if we use the identification \cite{math2} for Eq.(\ref{plucker}) and
Eq.(\ref{point})
\begin{align}
D^{\prime}  &  =z_{12}=ay-bx,\nonumber\\
-x_{21}  &  =z_{13}=az-cx,\nonumber\\
x_{11}  &  =z_{14}=aw-dx,\nonumber\\
-x_{22}  &  =z_{23}=bz-cy,\nonumber\\
x_{12}  &  =z_{24}=bw-dy,\nonumber\\
1  &  =z_{34}=cw-dz \label{identity}%
\end{align}
where%

\[
D^{\prime}=x_{11}x_{22}-x_{12}x_{21}%
\]
is fixed by the quadratic relations in Eq.(\ref{rel}), and restrict
$[z_{12}:z_{13}:z_{14}:z_{23}:z_{24}:z_{34}]$ to be a real line, we end up
with the correspondence%
\begin{equation}
\Delta^{\circledast}\Delta=\beta_{\lbrack a:b:c:d]}\alpha_{\lbrack x:y:z:w]}.
\label{id5}%
\end{equation}
So the jumping line connecting two points $[a:b:c:d]$ and $[x:y:z:w]$ on
$CP^{3}$ can be calculated from the jumping line equation%
\begin{equation}
\det\beta_{\lbrack a:b:c:d]}\alpha_{\lbrack x:y:z:w]}=0. \label{id1}%
\end{equation}
On the other hand, the corresponding singular point on $S^{4}$ associated with
jumping line can be calculated from Eq.(\ref{points}).

Before moving to the next subsection, let's look at the identification in
Eq.(\ref{identity}) in more details. Note that the four complex number
$(x_{11},x_{12},x_{21},x_{22})$ in Eq.(\ref{identity}) represent a line in
$CP^{3}$. If we choose $[a:b:c:d]=$ $\sigma\lbrack x:y:z:w]=[\bar{y}:-\bar
{x}:\bar{w}:-\bar{z}]$ in Eq.(\ref{identity}) and Eq.(\ref{id5}), we get%
\begin{align}
D^{\prime}  &  =z_{12}=\bar{y}y-(-\bar{x})x=\bar{y}y+\bar{x}x,\nonumber\\
-x_{21}  &  =z_{13}=\bar{y}z-\bar{w}x=\bar{y}z-\bar{w}x,\nonumber\\
x_{11}  &  =z_{14}=\bar{y}w-(-\bar{z})x=\bar{y}w+\bar{z}x,\nonumber\\
-x_{22}  &  =z_{23}=-\bar{x}z-\bar{w}y=-(\bar{x}z+\bar{w}y),\nonumber\\
x_{12}  &  =z_{24}=-\bar{x}w-(-\bar{z})y=-\bar{x}w+\bar{z}y,\nonumber\\
1  &  =z_{34}=\bar{w}w-(-\bar{z})z=\bar{w}w+\bar{z}z, \label{id2}%
\end{align}
and%
\begin{equation}
\Delta^{\circledast}\Delta=\beta_{\sigma\lbrack x:y:z:w]}\alpha_{\lbrack
x:y:z:w]}. \label{id6}%
\end{equation}
One can easily see that%
\begin{align}
x_{11}  &  =\bar{x}_{22},\nonumber\\
x_{12}  &  =-\bar{x}_{21},
\end{align}
which constrain $(x_{11},x_{12},x_{21},x_{22})$ to contain only four real
parameters to represent a real line over $S^{4}$. This real line is in an one
to one correspondence with a point $x$ with four real coordinates on $S^{4}$.
To be more specific, with the identification in Eq.(\ref{point}), the four
real coordinates in $x_{\mu}=(x_{0},x_{1},x_{2},x_{3})$ represents a point on
$S^{4}$, while $(x_{11},x_{12},x_{21},x_{22})$ in Eq.(\ref{id2}) represents
the corresponding real line in $CP^{3}$ over $S^{4}$. On the other hand,
Eq.(\ref{id6}) gives an exact relation between coordinates of the sheaf point
$[x:y:z:w]$ on $CP^{3}$ and coordinates of the corresponding singular point
$(x_{0},x_{1},x_{2},x_{3})$ on $S^{4}$.

To compare the parametrization used in Eq.(\ref{ii}), we note that
Eq.(\ref{ii}) can be further calculated to be%
\begin{align}
&  \pi(\left[  x:y:z:w\right]  )=\left[  x+ye_{2}:z+we_{2}\right] \nonumber\\
&  =\left[  x^{0}e_{0}+x^{1}e_{1}+y^{0}e_{2}+y^{1}e_{3}:z^{0}e_{0}+z^{1}%
e_{1}+w^{0}e_{2}+w^{1}e_{3}\right] \nonumber\\
&  \simeq\lbrack\left(  z^{0}e_{0}-z^{1}e_{1}-w^{0}e_{2}-w^{1}e_{3}\right)
\left(  x^{0}e_{0}+x^{1}e_{1}+y^{0}e_{2}+y^{1}e_{3}\right) \nonumber\\
&  :\left(  z^{0}e_{0}-z^{1}e_{1}-w^{0}e_{2}-w^{1}e_{3}\right)  (z^{0}%
e_{0}+z^{1}e_{1}+w^{0}e_{2}+w^{1}e_{3})]\nonumber\\
&  \simeq\left[  \frac{%
\begin{array}
[c]{c}%
\left(  z^{0}x^{0}+z^{1}x^{1}+w^{0}y^{0}+w^{1}y^{1}\right)  e_{0}+\left(
z^{0}x^{1}-z^{1}x^{0}-w^{0}y^{1}+w^{1}y^{0}\right)  e_{1}\\
+\left(  z^{0}y^{0}+z^{1}y^{1}-w^{0}x^{0}-w^{1}x^{1}\right)  e_{2}+\left(
z^{0}y^{1}-z^{1}y^{0}+w^{0}x^{1}-w^{1}x^{0}\right)  e_{3}%
\end{array}
}{\left(  \left(  z^{0}\right)  ^{2}+\left(  z^{1}\right)  ^{2}+\left(
w^{0}\right)  ^{2}+\left(  w^{1}\right)  ^{2}\right)  }:e_{0}\right]
\nonumber\\
&  =\left[  x_{0}e_{0}-x_{1}e_{3}+x_{2}e_{2}-x_{1}e_{3}:e_{0}\right]
\end{align}
where in the last step of the above calculation, we have used the
identifications in Eq.(\ref{point}) and Eq.(\ref{id2}). The quaternion
$(x_{0}e_{0}-x_{1}e_{3}+x_{2}e_{2}-x_{1}e_{3})$ in the above equation
represents a point in $S^{4}$ with parametrization used in Eq.(\ref{ii}) which
is different from parametrization used in Eq.(\ref{point}).

As an application of the above calculation, we can calculate for example the
real line corresponding to the sheaf point $[0:0:1:1]$ or the \textit{sheaf
line} in short obtained in the previous section\bigskip\ to be%
\begin{align}
\lbrack0  &  :0:1:1]\wedge\sigma\lbrack0:0:1:1]=[0:0:1:1]\wedge\lbrack
0:0:1:-1]\nonumber\\
&  =[0:0:0:0:0:-2]\simeq\lbrack0:0:0:0:0:1]. \label{line}%
\end{align}
For the sheaf point $[0:0:1:i]$, similar calculation gives%
\begin{align}
\lbrack0  &  :0:1:i]\wedge\sigma\lbrack0:0:1:i]=[0:0:1:i]\wedge\lbrack
0:0:1:-i]\nonumber\\
&  =[0:0:0:0:0:-2i]\simeq\lbrack0:0:0:0:0:1].
\end{align}
So all four sheaf points calculated in the previous section lie on the same
sheaf line. To calculate the projection of the sheaf point $[0:0:1:1]$ on
$CP^{3}$ down to $S^{4}$, we note from Eq.(\ref{id2}) and Eq.(\ref{line}) that%
\begin{equation}
(x_{11},x_{12},x_{21},x_{22})=(0,0,0,0)
\end{equation}
which means
\begin{equation}
x_{\mu}=(0,0,0,0) \label{0000}%
\end{equation}
by Eq.(\ref{point}). The projection of all other three sheaf points on
$CP^{3}$ down to $S^{4}$ is $x_{\mu}=(0,0,0,0)$ too. Here we note that $S^{4}$
contains two parts%
\begin{equation}
S^{4}=R^{4}\cup\{\infty\},
\end{equation}
or, in the language of quaternion projective space in Eq.(\ref{hp}),%
\begin{equation}
S^{4}\cong HP^{1}=[R^{4}:1]\cup\lbrack1:0].
\end{equation}

\subsection{Properties of the sheaf line as jumping line}

For the YM $2$-instanton data obtained in the last section%
\begin{align}
J_{1}  &  =%
\begin{pmatrix}
a & 0\\
0 & ia
\end{pmatrix}
,J_{2}=%
\begin{pmatrix}
0 & ia\\
a & 0
\end{pmatrix}
,\nonumber\\
B_{11}  &  =%
\begin{pmatrix}
\frac{-ia}{\sqrt{2}} & 0\\
0 & \frac{ia}{\sqrt{2}}%
\end{pmatrix}
,B_{21}=%
\begin{pmatrix}
0 & _{\frac{a}{\sqrt{2}}}\\
\frac{a}{\sqrt{2}} & 0
\end{pmatrix}
,\nonumber\\
B_{12}  &  =%
\begin{pmatrix}
0 & _{\frac{a}{\sqrt{2}}}\\
\frac{a}{\sqrt{2}} & 0
\end{pmatrix}
,B_{22}=%
\begin{pmatrix}
\frac{-ia}{\sqrt{2}} & 0\\
0 & \frac{ia}{\sqrt{2}}%
\end{pmatrix}
,\nonumber\\
I_{1}  &  =%
\begin{pmatrix}
0 & a\\
-ia & 0
\end{pmatrix}
,I_{2}=%
\begin{pmatrix}
-a & 0\\
0 & ia
\end{pmatrix}
, \label{data}%
\end{align}
we can calculate the singular points on $S^{4}$ associated with the jumping
line. The two delta matrices can be written as
\begin{equation}
\Delta=%
\begin{pmatrix}
ae_{0} & -ae_{1}\\
x+\frac{-i}{\sqrt{2}}ae_{0} & \frac{ia}{\sqrt{2}}e_{1}\\
\frac{ia}{\sqrt{2}}e_{1} & x+\frac{i}{\sqrt{2}}ae_{0}%
\end{pmatrix}
,\Delta^{\circledast}=%
\begin{pmatrix}
ae_{0} & x^{\dagger}+\frac{-i}{\sqrt{2}}ae_{0} & \frac{-ia}{\sqrt{2}}e_{1}\\
ae_{1} & \frac{-ia}{\sqrt{2}}e_{1} & x^{\dagger}+\frac{i}{\sqrt{2}}ae_{0}%
\end{pmatrix}
,
\end{equation}
and their product can be calculated to be%
\begin{align}
\Delta^{\circledast}\Delta &  =%
\begin{pmatrix}
ae_{0} & x^{\dagger}+\frac{-i}{\sqrt{2}}ae_{0} & \frac{-ia}{\sqrt{2}}e_{1}\\
ae_{1} & \frac{-ia}{\sqrt{2}}e_{1} & x^{\dagger}+\frac{i}{\sqrt{2}}ae_{0}%
\end{pmatrix}%
\begin{pmatrix}
ae_{0} & -ae_{1}\\
x+\frac{-i}{\sqrt{2}}ae_{0} & \frac{ia}{\sqrt{2}}e_{1}\\
\frac{ia}{\sqrt{2}}e_{1} & x+\frac{i}{\sqrt{2}}ae_{0}%
\end{pmatrix}
\nonumber\\
&  =%
\begin{pmatrix}
\left\vert x\right\vert ^{2}-\sqrt{2}iax_{0} & \sqrt{2}iax_{1}\\
\sqrt{2}iax_{1} & \left\vert x\right\vert ^{2}+\sqrt{2}iax_{0}%
\end{pmatrix}
,
\end{align}
which gives%
\begin{equation}
\det\Delta^{\circledast}\Delta=\left(  x_{0}^{2}+x_{1}^{2}+x_{2}^{2}+x_{3}%
^{2}\right)  ^{2}+2a^{2}\left(  x_{0}^{2}+x_{1}^{2}\right)  . \label{det}%
\end{equation}
\newline We conclude that%
\begin{equation}
\left(  x_{0}^{2}+x_{1}^{2}+x_{2}^{2}+x_{3}^{2}\right)  ^{2}+2a^{2}\left(
x_{0}^{2}+x_{1}^{2}\right)  =0 \label{00}%
\end{equation}
gives the singular locus on $S^{4}.$ One important observation is that for the
special singular point%
\begin{equation}
x_{\mu}=(0,0,0,0)
\end{equation}
associated with sheaf line calculated in Eq.(\ref{0000}), Eq.(\ref{det}) gives%
\begin{equation}
\det\Delta^{\circledast}\Delta_{sheaf}=0. \label{det0}%
\end{equation}
So this sheaf line is indeed a jumping line. This is a general statement.
Indeed, for the case of sheaf lines, either $\alpha$ is not injective or
$\beta$ is not surjective. If $\alpha$ is not injective, then $\beta\alpha$ is
not injective, which implies $\det\beta\alpha=0$ or Eq.(\ref{det0}). If
$\beta$ is not surjective, then $\beta\alpha$ is not surjective, which again
implies $\det\beta\alpha=0$ or Eq.(\ref{det0}). This completes the proof that
sheaf lines are special jumping lines. We thus have seen that the following
equation holds%
\begin{equation}
\text{\{lines in }CP^{3}\text{\}}\supset\text{\{real lines over }%
S^{4}\text{\}}\supset\text{\{jumping lines over }S^{4}\text{\}}\supset
\text{\{sheaf lines over }S^{4}\text{\}}. \label{ppp}%
\end{equation}

To identify sheaf lines among jumping lines, in the next section, we will see
that the order of singularity of the connection $A$ and the field strength $F$
at the singular point on $S^{4}$ associated with sheaf line in $CP^{3}$ is
higher than those of other singular points associated with normal jumping lines.

Another interesting observation is that the location of the sheaf point
$x_{\mu}=(0,0,0,0)$ seems reasonable since it is exactly the geometrical
center of "positions" $y_{11}$ and $y_{22}$ of the $2$-instantons in the ADHM
data \cite{Ann2}%
\begin{equation}
y_{11}=-de_{0}=%
\begin{bmatrix}
-d & 0\\
0 & -d
\end{bmatrix}
,y_{22}=de_{0}=%
\begin{bmatrix}
d & 0\\
0 & d
\end{bmatrix}
,
\end{equation}
which we have chosen to obtain Eq.(\ref{data1}) and Eq.(\ref{data2}).

For the case of diagonal CFTW $SL(2,C)$ $2$-instanton solutions, there are no
sheaf lines although the jumping lines do exist. The jumping lines or singular
locus calculated in Eq.(\ref{zero}) are%
\begin{equation}
P_{4}(x)=0,\text{ \ }P_{3}(x)=0.
\end{equation}

The result of Eq.(\ref{00}) can also be obtained by calculating the
determinant of $\beta\alpha$ in Eq.(\ref{id1})%
\begin{equation}
\det\beta\alpha=\left(  x_{0}^{2}+x_{1}^{2}+x_{2}^{2}+x_{3}^{2}\right)
^{2}+2a^{2}\left(  x_{0}^{2}+x_{1}^{2}\right)  , \label{detba}%
\end{equation}
\newline which is consistent with Eq.(\ref{det}). In this calculation, we have
used the identifications in Eq.(\ref{id2}) and Eq.(\ref{point}).

Finally, to understand the change of dimensionality of vector bundles at the
sheaf points, we can calculate the ranks of $\alpha$ and $\beta$ for a given
ADHM data at the sheaf points. For the ADHM data in Eq.(\ref{data}),
\begin{align}
\alpha_{\lbrack x:y:z:w]}  &  =%
\begin{pmatrix}
I_{2\times2}\\
0_{2\times2}\\
0_{2\times2}%
\end{pmatrix}
x+%
\begin{pmatrix}
0_{2\times2}\\
I_{2\times2}\\
0_{2\times2}%
\end{pmatrix}
y+%
\begin{pmatrix}
B_{11}\\
B_{12}\\
J_{1}%
\end{pmatrix}
z+%
\begin{pmatrix}
B_{21}\\
B_{22}\\
J_{2}%
\end{pmatrix}
w\nonumber\\
&  =%
\begin{pmatrix}
I_{2\times2}\\
0_{2\times2}\\
0_{2\times2}%
\end{pmatrix}
x+%
\begin{pmatrix}
0_{2\times2}\\
I_{2\times2}\\
0_{2\times2}%
\end{pmatrix}
y+%
\begin{pmatrix}
\frac{-ia}{\sqrt{2}} & 0\\
0 & \frac{ia}{\sqrt{2}}\\
0 & \frac{a}{\sqrt{2}}\\
\frac{a}{\sqrt{2}} & 0\\
a & 0\\
0 & ia
\end{pmatrix}
z+%
\begin{pmatrix}
0 & _{\frac{a}{\sqrt{2}}}\\
\frac{a}{\sqrt{2}} & 0\\
\frac{-ia}{\sqrt{2}} & 0\\
0 & \frac{ia}{\sqrt{2}}\\
0 & ia\\
a & 0
\end{pmatrix}
w
\end{align}
and%
\begin{align}
\beta_{\lbrack x:y:z:w]}  &  =%
\begin{pmatrix}
0_{2\times2} & I_{2\times2} & 0_{2\times2}%
\end{pmatrix}
x+%
\begin{pmatrix}
-I_{2\times2} & 0_{2\times2} & 0_{2\times2}%
\end{pmatrix}
y+%
\begin{pmatrix}
-B_{12} & B_{11} & I_{1}%
\end{pmatrix}
z+%
\begin{pmatrix}
-B_{22} & B_{21} & I_{2}%
\end{pmatrix}
w\nonumber\\
&  =%
\begin{pmatrix}
0_{2\times2} & I_{2\times2} & 0_{2\times2}%
\end{pmatrix}
x+%
\begin{pmatrix}
-I_{2\times2} & 0_{2\times2} & 0_{2\times2}%
\end{pmatrix}
y\nonumber\\
&  +%
\begin{pmatrix}
0 & -\frac{a}{\sqrt{2}} & \frac{-ia}{\sqrt{2}} & 0 & 0 & a\\
-\frac{a}{\sqrt{2}} & 0 & 0 & \frac{ia}{\sqrt{2}} & -ia & 0
\end{pmatrix}
z+%
\begin{pmatrix}
\frac{ia}{\sqrt{2}} & 0 & 0 & \frac{a}{\sqrt{2}} & -a & 0\\
0 & \frac{-ia}{\sqrt{2}} & \frac{a}{\sqrt{2}} & 0 & 0 & ia
\end{pmatrix}
w,
\end{align}
we can calculate $\alpha$ and $\beta$ at the sheaf point $[x:y:z:w]=[0:0:1:1]$
to be $(a\neq0)$%
\begin{equation}
\alpha_{\lbrack0:0:1:1]}=%
\begin{pmatrix}
\frac{-ia}{\sqrt{2}} & \frac{a}{\sqrt{2}}\\
\frac{a}{\sqrt{2}} & \frac{ia}{\sqrt{2}}\\
\frac{-ia}{\sqrt{2}} & \frac{a}{\sqrt{2}}\\
\frac{a}{\sqrt{2}} & \frac{ia}{\sqrt{2}}\\
a & ia\\
a & ia
\end{pmatrix}
,\beta_{\lbrack0:0:1:1]}=%
\begin{pmatrix}
\frac{ia}{\sqrt{2}} & -\frac{a}{\sqrt{2}} & \frac{-ia}{\sqrt{2}} & \frac
{a}{\sqrt{2}} & -a & a\\
-\frac{a}{\sqrt{2}} & \frac{-ia}{\sqrt{2}} & \frac{a}{\sqrt{2}} & \frac
{ia}{\sqrt{2}} & -ia & ia
\end{pmatrix}
,
\end{equation}
which are both of rank $1$. So the dimensions of Ker$\alpha_{\lbrack0:0:1:1]}$
and Ker$\beta_{\lbrack0:0:1:1]}$ are $1$ and $5$ respectively, which imply the
dimension of the quotient space%
\begin{equation}
\dim(\text{Ker}\beta_{\lbrack0:0:1:1]}/\operatorname{Im}\alpha_{\lbrack
0:0:1:1]})=5-1=4. \label{5}%
\end{equation}
Note that for points other than sheaf points $\dim($Ker$\beta
/\operatorname{Im}\alpha)=4-2=2$.

Similarly, $\alpha$ and $\beta$ at point $[x:y:z:w]=[0:0:1:-1]$ are%
\begin{equation}
\alpha_{\lbrack0:0:1:-1]}=%
\begin{pmatrix}
\frac{-ia}{\sqrt{2}} & -\frac{a}{\sqrt{2}}\\
-\frac{a}{\sqrt{2}} & \frac{ia}{\sqrt{2}}\\
\frac{ia}{\sqrt{2}} & \frac{a}{\sqrt{2}}\\
\frac{a}{\sqrt{2}} & \frac{-ia}{\sqrt{2}}\\
a & -ia\\
-a & ia
\end{pmatrix}
,\beta_{\lbrack0:0:1:-1]}=%
\begin{pmatrix}
-\frac{ia}{\sqrt{2}} & -\frac{a}{\sqrt{2}} & \frac{-ia}{\sqrt{2}} & -\frac
{a}{\sqrt{2}} & a & a\\
-\frac{a}{\sqrt{2}} & \frac{ia}{\sqrt{2}} & -\frac{a}{\sqrt{2}} & \frac
{ia}{\sqrt{2}} & -ia & -ia
\end{pmatrix}
,
\end{equation}
which are both of rank $1$, and the dimension of the quotient space is $4$,
same with Eq.(\ref{5}).

Similarly, one can calculate $\alpha$ and $\beta$ with ADHM data in
Eq.(\ref{data2}) at the sheaf point $[x:y:z:w]=[0:0:1:i]$ to be%
\begin{equation}
\alpha_{\lbrack0:0:1:i]}=%
\begin{pmatrix}
\frac{-ia}{\sqrt{2}} & \frac{a}{\sqrt{2}}\\
\frac{a}{\sqrt{2}} & \frac{ia}{\sqrt{2}}\\
\frac{a}{\sqrt{2}} & \frac{ia}{\sqrt{2}}\\
\frac{ia}{\sqrt{2}} & -\frac{a}{\sqrt{2}}\\
a & ia\\
ia & -a
\end{pmatrix}
,\beta_{\lbrack0:0:1:i]}=\
\begin{pmatrix}
-\frac{a}{\sqrt{2}} & -\frac{ia}{\sqrt{2}} & \frac{-ia}{\sqrt{2}} & \frac
{a}{\sqrt{2}} & -ia & a\\
-\frac{ia}{\sqrt{2}} & \frac{a}{\sqrt{2}} & \frac{a}{\sqrt{2}} & \frac
{ia}{\sqrt{2}} & a & ia
\end{pmatrix}
,
\end{equation}
and at the sheaf point $[x:y:z:w]=[0:0:1:-i]$ to be%
\begin{equation}
\alpha_{\lbrack0:0:1:-i]}=%
\begin{pmatrix}
\frac{-ia}{\sqrt{2}} & \frac{-a}{\sqrt{2}}\\
\frac{-a}{\sqrt{2}} & \frac{ia}{\sqrt{2}}\\
\frac{-a}{\sqrt{2}} & \frac{ia}{\sqrt{2}}\\
\frac{ia}{\sqrt{2}} & \frac{a}{\sqrt{2}}\\
a & -ia\\
-ia & -a
\end{pmatrix}
,\beta_{\lbrack0:0:1:-i]}=%
\begin{pmatrix}
\frac{a}{\sqrt{2}} & \frac{-ia}{\sqrt{2}} & \frac{-ia}{\sqrt{2}} & \frac
{-a}{\sqrt{2}} & ia & a\\
-\frac{ia}{\sqrt{2}} & \frac{-a}{\sqrt{2}} & \frac{-a}{\sqrt{2}} & \frac
{ia}{\sqrt{2}} & a & -ia
\end{pmatrix}
.
\end{equation}
We can also calculate $\alpha$ and $\beta$ with ADHM data in Eq.(\ref{data3})
at the sheaf point $[x:y:z:w]=[0:0:1:-1]$ to be%
\begin{equation}
\alpha_{\left[  0:0:1:-1\right]  }^{\prime}=%
\begin{pmatrix}
\frac{i\bar{a}}{\sqrt{2}} & \frac{\bar{a}}{\sqrt{2}}\\
\frac{\bar{a}}{\sqrt{2}} & \frac{-i\bar{a}}{\sqrt{2}}\\
-\frac{i\bar{a}}{\sqrt{2}} & -\frac{\bar{a}}{\sqrt{2}}\\
-\frac{\bar{a}}{\sqrt{2}} & \frac{i\bar{a}}{\sqrt{2}}\\
\bar{a} & -i\bar{a}\\
-\bar{a} & i\bar{a}%
\end{pmatrix}
,\beta_{\left[  0:0:1:-1\right]  }^{\prime}=%
\begin{pmatrix}
\frac{i\bar{a}}{\sqrt{2}} & \frac{\bar{a}}{\sqrt{2}} & \frac{i\bar{a}}%
{\sqrt{2}} & \frac{\bar{a}}{\sqrt{2}} & \bar{a} & \bar{a}\\
\frac{\bar{a}}{\sqrt{2}} & \frac{-i\bar{a}}{\sqrt{2}} & \frac{\bar{a}}%
{\sqrt{2}} & \frac{-i\bar{a}}{\sqrt{2}} & -i\bar{a} & -i\bar{a}%
\end{pmatrix}
.
\end{equation}
In all cases of sheaf points, we find that $\dim($Ker$\beta/\operatorname{Im}%
\alpha)=5-1=4.$

\section{Singularity structure of $A$ and $F$ associated with sheaf line}

In the previous section, we showed that all sheaf lines are jumping lines.
What makes sheaf lines different from the normal jumping lines on $S^{4}$
spacetime? In this section, we will show that the order of singularity of the
connection $A$ and the field strength $F$ at the singular point on $S^{4}$
associated with sheaf line in $CP^{3}$ is higher than those of other singular
points associated with normal jumping lines.

\subsection{Singularity structure of connection}

In the explicit calculation of $SU(2)$ instanton connections, one needs to do
a large gauge transformation to remove all the singularities on $S^{4}$. This
can be easily done for the case of $1$-instanton. For the case of diagonal
CFTW $2$-instanton, see the choice of large gauge transformation function in
\cite{1977}. For the $SL(2,C)$ instanton connections, in addition to the
removable singularities, there exist non-removable singularities \cite{Ann}
associated with jumping lines in $CP^{3}$. For example, for the case of
$SL(2,C)$ diagonal CFTW $2$-instanton, these non-removable singularities can
be calculated from Eq.(\ref{zero}).

For the non-diagonal ADHM $2$-instanton sheaves of the present case, we will
use similar technique and identify only non-removable singularities which
containing the singularity structure associated with the sheaf line. The
explicit form of the $2$-instanton connection without removable singularities
will not be calculated. However, it is interesting to see that the explicit
form of $2$-instanton field strength without removable singularities can be
exactly calculated and will be given in the next subsection. We begin with the
two delta matrices with ADHM data given in Eq.(\ref{data})%

\begin{align}
\Delta &  =%
\begin{pmatrix}
ae_{0} & -ae_{1}\\
x+\frac{-i}{\sqrt{2}}ae_{0} & \frac{ia}{\sqrt{2}}e_{1}\\
\frac{ia}{\sqrt{2}}e_{1} & x+\frac{i}{\sqrt{2}}ae_{0}%
\end{pmatrix}
,\\
\Delta^{\circledast}  &  =%
\begin{pmatrix}
ae_{0} & x^{\dagger}+\frac{-i}{\sqrt{2}}ae_{0} & \frac{-ia}{\sqrt{2}}e_{1}\\
ae_{1} & \frac{-ia}{\sqrt{2}}e_{1} & x^{\dagger}+\frac{i}{\sqrt{2}}ae_{0}%
\end{pmatrix}
. \label{tt}%
\end{align}
To calculate the connection, we need to first identify $v$ vector which
satisfies $\Delta^{\circledast}v=0$ or%
\begin{equation}%
\begin{pmatrix}
ae_{0} & x^{\dagger}+\frac{-i}{\sqrt{2}}ae_{0} & \frac{-ia}{\sqrt{2}}e_{1}\\
ae_{1} & \frac{-ia}{\sqrt{2}}e_{1} & x^{\dagger}+\frac{i}{\sqrt{2}}ae_{0}%
\end{pmatrix}%
\begin{pmatrix}
v_{1}\\
v_{2}\\
v_{3}%
\end{pmatrix}
=0 \label{ttt}%
\end{equation}
which means%
\begin{align}
ae_{0}v_{1}+\left(  x^{\dagger}+\frac{-i}{\sqrt{2}}ae_{0}\right)
v_{2}+\left(  \frac{-ia}{\sqrt{2}}e_{1}\right)  v_{3}  &  =0,\\
ae_{1}v_{1}+\left(  \frac{-ia}{\sqrt{2}}e_{1}\right)  v_{2}+\left(
x^{\dagger}+\frac{i}{\sqrt{2}}ae_{0}\right)  v_{3}  &  =0,
\end{align}
from which one can solve $v_{2}$ and $v_{1}$ to be%
\begin{equation}
v_{2}=\frac{-xe_{1}x^{\dagger}}{\left\vert x\right\vert ^{2}}v_{3},
\end{equation}%
\begin{equation}
v_{1}=\frac{1}{a}\left[  e_{1}x^{\dagger}+\frac{ia}{\sqrt{2}}\left(
e_{1}-\frac{xe_{1}x^{\dagger}}{\left\vert x\right\vert ^{2}}\right)  \right]
v_{3}.
\end{equation}
Finally $v$ and $v^{\circledast}$ can be written as%
\begin{equation}
v=%
\begin{pmatrix}
v_{1}\\
v_{2}\\
v_{3}%
\end{pmatrix}
=%
\begin{pmatrix}
\frac{1}{a}\left[  e_{1}x^{\dagger}+\frac{ia}{\sqrt{2}}\left(  e_{1}%
-\frac{xe_{1}x^{\dagger}}{\left\vert x\right\vert ^{2}}\right)  \right]
v_{3}\\
\frac{-xe_{1}x^{\dagger}}{\left\vert x\right\vert ^{2}}v_{3}\\
v_{3}%
\end{pmatrix}
\end{equation}
and%
\begin{equation}
v^{\circledast}=%
\begin{pmatrix}
v_{3}^{\circledast}\frac{1}{a}\left[  -xe_{1}+\frac{ia}{\sqrt{2}}\left(
-e_{1}+\frac{xe_{1}x^{\dagger}}{\left\vert x\right\vert ^{2}}\right)  \right]
, & v_{3}^{\circledast}\frac{xe_{1}x^{\dagger}}{\left\vert x\right\vert ^{2}%
}, & v_{3}^{\circledast}%
\end{pmatrix}
\end{equation}
respectively. The next step is to do the normalization condition%
\begin{equation}
v^{\circledast}v=1
\end{equation}
or%
\begin{equation}
v_{3}^{\circledast}\left\{
\begin{array}
[c]{c}%
\frac{1}{a^{2}}\left[  -xe_{1}+\frac{ia}{\sqrt{2}}\left(  -e_{1}+\frac
{xe_{1}x^{\dagger}}{\left\vert x\right\vert ^{2}}\right)  \right]  \left[
e_{1}x^{\dagger}+\frac{ia}{\sqrt{2}}\left(  e_{1}-\frac{xe_{1}x^{\dagger}%
}{\left\vert x\right\vert ^{2}}\right)  \right] \\
+\frac{xe_{1}x^{\dagger}}{\left\vert x\right\vert ^{2}}\left(  \frac
{-xe_{1}x^{\dagger}}{\left\vert x\right\vert ^{2}}\right)  +1
\end{array}
\right\}  v_{3}=1
\end{equation}
to extract the non-removable singular factor similar to Eq.(\ref{zero}). After
some lengthy calculation, we end up with
\begin{equation}
v_{3}^{\circledast}\left\{  \frac{1}{a^{2}}\left[  \left(  \frac{\left\vert
x\right\vert ^{4}+2a^{2}\left(  x_{0}^{2}+x_{1}^{2}\right)  }{\left\vert
x\right\vert ^{2}}\right)  \right]  \right\}  v_{3}=1
\end{equation}
where $\left\vert x\right\vert ^{4}=\left(  x_{0}^{2}+x_{1}^{2}+x_{2}%
^{2}+x_{3}^{2}\right)  ^{2}$. So the normalization can be done by setting%
\begin{equation}
v_{3}=\frac{a\left\vert x\right\vert }{\sqrt{\left\vert x\right\vert
^{4}+2a^{2}\left(  x_{0}^{2}+x_{1}^{2}\right)  }}, \label{33}%
\end{equation}
and the normalized $\nu$ and $\nu^{\circledast}$ vector can be written as%
\begin{align}
v  &  =%
\begin{pmatrix}
v_{1}\\
v_{2}\\
v_{3}%
\end{pmatrix}
=\frac{a\left\vert x\right\vert }{\sqrt{\left\vert x\right\vert ^{4}%
+2a^{2}\left(  x_{0}^{2}+x_{1}^{2}\right)  }}%
\begin{pmatrix}
\frac{1}{a}\left[  e_{1}x^{\dagger}+\frac{ia}{\sqrt{2}}\left(  e_{1}%
-\frac{xe_{1}x^{\dagger}}{\left\vert x\right\vert ^{2}}\right)  \right] \\
\frac{-xe_{1}x^{\dagger}}{\left\vert x\right\vert ^{2}}\\
1
\end{pmatrix}
,\label{v1}\\
v^{\circledast}  &  =\frac{a\left\vert x\right\vert }{\sqrt{\left\vert
x\right\vert ^{4}+2a^{2}\left(  x_{0}^{2}+x_{1}^{2}\right)  }}%
\begin{pmatrix}
\frac{1}{a}\left[  -xe_{1}+\frac{ia}{\sqrt{2}}\left(  -e_{1}+\frac
{xe_{1}x^{\dagger}}{\left\vert x\right\vert ^{2}}\right)  \right]  , &
\frac{xe_{1}x^{\dagger}}{\left\vert x\right\vert ^{2}}, & 1
\end{pmatrix}
. \label{v2}%
\end{align}
The connection $A$ can be written as%
\begin{equation}
A_{\mu}=v^{\circledast}\partial_{\mu}v.
\end{equation}

It turns out that in order to extract non-removable singularity structure of
$A$, one needs only check the normalization factor calculated in
Eq.(\ref{33}). This is similar to the calculation in Eq.(\ref{zero}) for the
case of $SL(2,C)$ CFTW $2$-instanton. The factor inside the square root in
Eq.(\ref{33}) is exactly the same with $\det\Delta^{\circledast}\Delta$ and
$\det\beta\alpha$ calculated in Eq.(\ref{det}) and Eq.(\ref{detba})
respectively. We conclude that the non-removable singularities of the
connection $A$ occur at%
\begin{equation}
\left\vert x\right\vert ^{4}+2a^{2}\left(  x_{0}^{2}+x_{1}^{2}\right)  =0,
\end{equation}
which is the same with the singular locus calculated in Eq.(\ref{00}).

\subsection{Singularity structure of field strength}

In this subsection, we go one step further to calculate the singularity
structure of field strength $F.$ It turns out that $F$ without removable
singularities is much more easier to calculate than $A$. The formula for the
field strength of $SL(2,C)$ ADHM instanton calculated in \cite{Ann} was
\begin{equation}
F_{\mu\nu}=v^{\circledast}b\left(  e_{\mu}e_{\nu}^{\dagger}-e_{\nu}e_{\mu
}^{\dagger}\right)  fb^{\circledast}v \label{f2}%
\end{equation}
where $v$ and $v^{\circledast}$ were given in Eq.(\ref{v1}) and Eq.(\ref{v2})
respectively, and other factors can be calculated to be
\begin{align}
\Delta^{\circledast}\Delta &  =f^{-1}=%
\begin{pmatrix}
\left\vert x\right\vert ^{2}-\sqrt{2}iax_{0} & \sqrt{2}iax_{1}\\
\sqrt{2}iax_{1} & \left\vert x\right\vert ^{2}+\sqrt{2}iax_{0}%
\end{pmatrix}
,\label{f3}\\
f  &  =\frac{1}{\left\vert x\right\vert ^{4}+2a^{2}\left(  x_{0}^{2}+x_{1}%
^{2}\right)  }%
\begin{pmatrix}
\left\vert x\right\vert ^{2}+\sqrt{2}iax_{0} & -\sqrt{2}iax_{1}\\
-\sqrt{2}iax_{1} & \left\vert x\right\vert ^{2}-\sqrt{2}iax_{0}%
\end{pmatrix}
,\label{f4}\\
b  &  =%
\begin{pmatrix}
0 & 0\\
1 & 0\\
0 & 1
\end{pmatrix}
,b^{\circledast}=%
\begin{pmatrix}
0 & 1 & 0\\
0 & 0 & 1
\end{pmatrix}
. \label{f5}%
\end{align}
The field strength can then be calculated to be%
\begin{align}
F_{\mu\nu}  &  =\frac{a\left\vert x\right\vert }{\sqrt{\left\vert x\right\vert
^{4}+2a^{2}\left(  x_{0}^{2}+x_{1}^{2}\right)  }}%
\begin{pmatrix}
\frac{1}{a}\left[  -xe_{1}+\frac{ia}{\sqrt{2}}\left(  -e_{1}+\frac
{xe_{1}x^{\dagger}}{\left\vert x\right\vert ^{2}}\right)  \right]  , &
\frac{xe_{1}x^{\dagger}}{\left\vert x\right\vert ^{2}}, & 1
\end{pmatrix}
\nonumber\\
&  \cdot%
\begin{pmatrix}
0 & 0\\
1 & 0\\
0 & 1
\end{pmatrix}
\left(  e_{\mu}e_{\nu}^{\dagger}-e_{\nu}e_{\mu}^{\dagger}\right)  \frac{%
\begin{pmatrix}
\left\vert x\right\vert ^{2}+\sqrt{2}iax_{0} & -\sqrt{2}iax_{1}\\
-\sqrt{2}iax_{1} & \left\vert x\right\vert ^{2}-\sqrt{2}iax_{0}%
\end{pmatrix}
}{\left\vert x\right\vert ^{4}+2a^{2}\left(  x_{0}^{2}+x_{1}^{2}\right)
}\nonumber\\
&  \cdot%
\begin{pmatrix}
0 & 1 & 0\\
0 & 0 & 1
\end{pmatrix}
\frac{a\left\vert x\right\vert }{\sqrt{\left\vert x\right\vert ^{4}%
+2a^{2}\left(  x_{0}^{2}+x_{1}^{2}\right)  }}%
\begin{pmatrix}
\frac{1}{a}\left[  e_{1}x^{\dagger}+\frac{ia}{\sqrt{2}}\left(  e_{1}%
-\frac{xe_{1}x^{\dagger}}{\left\vert x\right\vert ^{2}}\right)  \right] \\
\frac{-xe_{1}x^{\dagger}}{\left\vert x\right\vert ^{2}}\\
1
\end{pmatrix}
\nonumber\\
&  =\frac{a^{2}}{\left[  \left\vert x\right\vert ^{4}+2a^{2}\left(  x_{0}%
^{2}+x_{1}^{2}\right)  \right]  ^{2}}%
\begin{pmatrix}
\frac{xe_{1}x^{\dagger}}{\left\vert x\right\vert }, & \left\vert x\right\vert
\end{pmatrix}
\left(  e_{\mu}e_{\nu}^{\dagger}-e_{\nu}e_{\mu}^{\dagger}\right) \nonumber\\
&  \cdot%
\begin{pmatrix}
\left\vert x\right\vert ^{2}+\sqrt{2}iax_{0} & -\sqrt{2}iax_{1}\\
-\sqrt{2}iax_{1} & \left\vert x\right\vert ^{2}-\sqrt{2}iax_{0}%
\end{pmatrix}%
\begin{pmatrix}
\frac{-xe_{1}x^{\dagger}}{\left\vert x\right\vert }\\
\left\vert x\right\vert
\end{pmatrix}
. \label{field}%
\end{align}

It is important to see that there are non-removable singularities in the
prefactor of Eq.(\ref{field}). In addition, removable singularity shows up in
$%
\begin{pmatrix}
\frac{xe_{1}x^{\dagger}}{\left\vert x\right\vert }, & \left\vert x\right\vert
\end{pmatrix}
$ and $%
\begin{pmatrix}
\frac{xe_{1}x^{\dagger}}{\left\vert x\right\vert }, & \left\vert x\right\vert
\end{pmatrix}
,$ which surprisingly can be gauged away by preforming a large gauge
transformation with simple gauge function in the quaternion form as following%
\begin{align}
F_{\mu\nu}^{\prime}  &  =\frac{x^{\dagger}}{\left\vert x\right\vert }F_{\mu
\nu}\frac{x}{\left\vert x\right\vert }\nonumber\\
&  =\frac{a^{2}}{\left[  \left\vert x\right\vert ^{4}+2a^{2}\left(  x_{0}%
^{2}+x_{1}^{2}\right)  \right]  ^{2}}%
\begin{pmatrix}
e_{1}x^{\dagger}, & x^{\dagger}%
\end{pmatrix}
\left(  e_{\mu}e_{\nu}^{\dagger}-e_{\nu}e_{\mu}^{\dagger}\right) \nonumber\\
&  \cdot%
\begin{pmatrix}
\left\vert x\right\vert ^{2}+\sqrt{2}iax_{0} & -\sqrt{2}iax_{1}\\
-\sqrt{2}iax_{1} & \left\vert x\right\vert ^{2}-\sqrt{2}iax_{0}%
\end{pmatrix}%
\begin{pmatrix}
-xe_{1}\\
x
\end{pmatrix}
\nonumber\\
&  =\frac{a^{2}}{\left[  \left\vert x\right\vert ^{4}+2a^{2}\left(  x_{0}%
^{2}+x_{1}^{2}\right)  \right]  ^{2}}%
\begin{pmatrix}
e_{1}, & 1
\end{pmatrix}
x^{\dagger}\left(  e_{\mu}e_{\nu}^{\dagger}-e_{\nu}e_{\mu}^{\dagger}\right)
x\nonumber\\
&  \cdot%
\begin{pmatrix}
\left\vert x\right\vert ^{2}+\sqrt{2}iax_{0} & -\sqrt{2}iax_{1}\\
-\sqrt{2}iax_{1} & \left\vert x\right\vert ^{2}-\sqrt{2}iax_{0}%
\end{pmatrix}%
\begin{pmatrix}
-e_{1}\\
1
\end{pmatrix}
. \label{fff}%
\end{align}
We can see that the non-removable singular points occur in $\left\vert
x\right\vert ^{4}+2a^{2}\left(  x_{0}^{2}+x_{1}^{2}\right)  =0$, which is
consistent with all our previous calculations.

It is interesting to see that the explicit form of $SL(2,C)$ YM $2$-instanton
field strength without removable singularities presented in Eq.(\ref{fff}) can
be exactly calculated! To the knowledge of the authors, it seems to be a very
difficult task, though it might not be impossible, to exactly calculate
$SU(2)$ YM $2$-instanton field strength with all singularities removed by a
suitable large gauge transformation. See the discussion for the choice of
large gauge transformation function in \cite{1977} for the case of $SU(2)$
CFTW $2$-instanton.

To be more precisely, if one uses the $SL(2,C)$ ADHM data calculated from the
costable condition of sheaf structure in Eq.(\ref{a}) to Eq.(\ref{c}), and
plugs this $SL(2,C)$ \textit{sheaf} ADHM data into $\Delta^{\circledast}$ in
Eq.(\ref{tt}), then the calculation of $\nu$ in Eq.(\ref{ttt}) and thus the
field strength $F$ in Eq.(\ref{f2}) will be greatly simplified. A closer look
for this solvability or simplification seems worthwhile.

Presumably, the simplification for the calculation of $SL(2,C)$ YM
$2$-instanton field strength is also due to the existence of the sheaf line
with associated one single singular point at $x=(0,0,0,0)$ on $S^{4}$, instead
of two removable singular points corresponding to two positions of $SU(2)$ YM
$2$-instanton before doing a large gauge transformation \cite{1977}.

\subsection{Order of singularity at the sheaf line}

In the paragraph after Eq.(\ref{det0}), we have shown that all sheaf lines are
special jumping lines. In this subsection we will first define the order of
singularity of a jumping line including a sheaf line. We will then give a
general prescription to calculate it. Recall that in the $SL(2,C)$ ADHM
construction, the jumping lines were defined by Eq.(\ref{ff}) and
Eq.(\ref{points}) which we reproduce in the following%
\begin{equation}
\Delta(x)^{\circledast}\Delta(x)=f^{-1}, \label{ffss}%
\end{equation}%
\begin{equation}
\det\Delta(x)^{\circledast}\Delta(x)=0. \label{pointss}%
\end{equation}

Note that there are no jumping lines for the $SU(2)$ YM instanton. For a given
ADHM data, the field strength can be calculated to be (see the example given
in Eq.(\ref{f2}) to Eq.(\ref{field}))
\begin{equation}
F_{\mu\nu}=v^{\circledast}(x)b(e_{\mu}e_{\nu}^{\dagger}-e_{\nu}e_{\mu
}^{\dagger})fb^{\circledast}v(x). \label{f2ss}%
\end{equation}
In the case of $SU(2)$, $v(x)$ (but not $f$) in general contains "removable
singularities" which can be gauged away by doing a "large gauge
transformation" $g$ \cite{1977}
\begin{align}
F_{\mu\nu}^{\prime}  &  =v^{\prime\circledast}(x)b(e_{\mu}e_{\nu}^{\dagger
}-e_{\nu}e_{\mu}^{\dagger})fb^{\circledast}v^{\prime}(x),\nonumber\\
&  =g^{\circledast}(x)v^{\circledast}(x)b(e_{\mu}e_{\nu}^{\dagger}-e_{\nu
}e_{\mu}^{\dagger})fb^{\circledast}v(x)g(x).
\end{align}

For the case of $SL(2,C)$ YM instantons, in addition to the "removable
singularities" in $v(x)$, $f$ contains "non-removable singularities" which can
not be gauged away and remain \cite{Ann}. We define the order of singularity
of a jumping line to be the singularity in $f$
\begin{equation}
f=(f^{-1})^{-1}=\frac{[Cof\left(  f^{-1}\right)  ]^{t}}{\det f^{-1}}%
=\frac{[Cof\left(  f^{-1}\right)  ]^{t}}{\det\Delta(x)^{\circledast}\Delta(x)}
\label{f8}%
\end{equation}
where $Cof$ means cofactor of a matrix. In the following we review \cite{Ann}
some explicit calculations of \bigskip$\det\Delta(x)^{\circledast}\Delta(x):$

\subsubsection{The geometry of $1$-instanton jumping lines}

The complete jumping lines of the $SL(2,C)$ $1$-instanton can be described by
ADHM data with $10$ parameters $y_{\mu}=p_{\mu}+iq_{\mu}$ and $\lambda$. To
study these singularities, let the real part of $\lambda^{2}$ be $c$ and
imaginary part of $\lambda^{2}$ be $d,$ we see that \cite{Ann}%

\begin{align}
\det\Delta(x)^{\circledast}\Delta(x)=  &  (|x-(p+qi)|_{c}^{2}+\lambda
^{2})=P_{2}(x)+iP_{1}(x)\nonumber\\
&  =[(x_{0}-p_{0})^{2}+(x_{1}-p_{1})^{2}+(x_{2}-p_{2})^{2}+(x_{3}-p_{3}%
)^{2}\nonumber\\
&  -(q_{0}^{2}+q_{1}^{2}+q_{2}^{2}+q_{3}^{2})]+c\nonumber\\
&  -2i[(x_{0}-p_{0})q_{0}+(x_{1}-p_{1})q_{1}+(x_{2}-p_{2})q_{2}+(x_{3}%
-p_{3})q_{3}-\frac{d}{2}]=0, \label{jump1}%
\end{align}
which implies%
\begin{align}
(x_{0}-p_{0})^{2}+(x_{1}-p_{1})^{2}+(x_{2}-p_{2})^{2}+(x_{3}-p_{3})^{2}  &
=(q_{0}^{2}+q_{1}^{2}+q_{2}^{2}+q_{3}^{2})-c,\label{p1}\\
(x_{0}-p_{0})q_{0}+(x_{1}-p_{1})q_{1}+(x_{2}-p_{2})q_{2}+(x_{3}-p_{3})q_{3}
&  =\frac{d}{2} \label{p2}%
\end{align}
where $P_{2}(x)$ and $P_{1}(x)$ are polynomials of $4$ variables with degree
$2$ and $1$ respectively. The geometry of the above singular structure was
discussed in details in \cite{Ann}. There is no sheaf line for $SL(2,C)$ $1$-instanton.

\subsubsection{The complete $2$-instanton and $3$-instanton jumping lines}

Since the complete $2$-instanton and $3$-instanton ADHM data were worked out
in \cite{CSW,Ann}, the explicit form of \bigskip$\det\Delta(x)^{\circledast
}\Delta(x)$ can be explicitly calculated and the corresponding jumping lines
can in principal be identified \cite{Ann}. Since the form of the $3$-instanton
case is very lengthy, we list as an example only the $2$-instanton case in the
following \cite{Ann}%
\begin{align}
\det\Delta_{2-ins}(x)^{\circledast}\Delta_{2-ins}(x)  &  =|x-y_{1}|_{c}%
^{2}|x-y_{2}|_{c}^{2}+|\lambda_{2}|_{c}^{2}|x-y_{1}|_{c}^{2}+|\lambda_{1}%
|_{c}^{2}|x-y_{2}|_{c}^{2}\nonumber\\
&  +y_{12}^{\circledast}(x-y_{1})y_{12}^{\circledast}(x-y_{2})+(x-y_{2}%
)^{\circledast}y_{12}(x-y_{1})^{\circledast}y_{12}\nonumber\\
&  -y_{12}^{\circledast}(x-y_{1})\lambda_{1}^{\circledast}\lambda_{2}%
-\lambda_{2}^{\circledast}\lambda_{1}(x-y_{1})^{\circledast}y_{12}\nonumber\\
&  -(x-y_{2})^{\circledast}y_{12}\lambda_{1}^{\circledast}\lambda_{2}%
-\lambda_{2}^{\circledast}\lambda_{1}y_{12}^{\circledast}(x-y_{2})\nonumber\\
&  +|y_{12}|_{c}^{2}(|\lambda_{2}|_{c}^{2}+|\lambda_{1}|_{c}^{2})+|y_{12}%
|_{c}^{4}. \label{4 degree}%
\end{align}
One sees that Eq.(\ref{4 degree}) is a polynomial of degree $4$ in $x.$ So the
order of singularity in $f$ is at most $4$ for a given ADHM data. In general,
the order of singularity in $f$ is at most $2k$ for a given $k$-instanton ADHM
data. Although the complete $2$-instanton jumping lines have been calculated
in Eq.(\ref{4 degree}), the existence of a special sheaf line was not known in
\cite{Ann}. One explicit example of a $2$-instanton sheaf line with order of
singularity $2$ was calculated in Eq.(\ref{f4}). We will discuss this example
in details later.

\subsubsection{A class of $k$-instanton jumping lines}

A class of $SL(2,C)$ $k$-instanton jumping lines, the $SL(2,C)$ CFTW or the
generalized $(M,N)$ instanton jumping lines were calculated to be zeros of the
following determinant\bigskip\ \cite{Ann}.%
\begin{equation}
\det\Delta(x)^{\circledast}\Delta(x)=|x-y_{1}|_{c}^{2}|x-y_{2}|_{c}^{2}%
\cdot\cdot\cdot|x-y_{k}|_{c}^{2}\phi=P_{2k}(x)+iP_{2k-1}(x) \label{pp}%
\end{equation}
where%
\begin{equation}
\phi=1+\frac{\lambda_{1}\lambda_{1}^{\circledast}}{|x-y_{1}|_{c}^{2}%
}+...+\frac{\lambda_{k}\lambda_{k}^{\circledast}}{|x-y_{k}|_{c}^{2}}.
\end{equation}
In Eq.(\ref{pp}), $P_{2k}(x)$ and $P_{2k-1}(x)$ are polynomials with degrees
$2k$ and $2k-1$ respectively. The case of $2$-instanton jumping lines was
calculated in Eq.(\ref{zero}). Unfortunately, it was shown \cite{Ann} that
there existed no sheaf lines for this case.

\subsubsection{Order of singularities at the sheaf line and jumping line}

In this subsection, we will show that the order of singularity calculated in
the previous subsections for connection and field strength at the singular
point $x_{\mu}=(0,0,0,0)$ on $S^{4}$ associated with the sheaf line in
$CP^{3}$ is higher than those of other singular points associated with normal
jumping lines. The function we want to study is in the denominator of the
prefactor in Eq.(\ref{fff})
\begin{align}
h(x_{0},x_{1},x_{2},x_{3})  &  =\left\vert x\right\vert ^{4}+2a^{2}\left(
x_{0}^{2}+x_{1}^{2}\right) \nonumber\\
&  =\left(  x_{0}^{2}+x_{1}^{2}+x_{2}^{2}+x_{3}^{2}\right)  ^{2}+2a^{2}\left(
x_{0}^{2}+x_{1}^{2}\right)  ,\text{ }a\in C,\text{ }a\neq0.
\end{align}
One can easily see that%
\begin{equation}
h(x_{0},x_{1},x_{2},x_{3})=0\text{ \ and \ }\partial_{\mu}h(x_{0},x_{1}%
,x_{2},x_{3})=0\text{ \ for }x_{\mu}=(0,0,0,0). \label{order}%
\end{equation}

We want to show that there is no spacetime point other than $x_{\mu
}=(0,0,0,0)$ which shares the same property as in Eq.(\ref{order}). This means
that we are looking for \textit{non-zero} solution for the following system of
equations%
\begin{equation}
\left(  x_{0}^{2}+x_{1}^{2}+x_{2}^{2}+x_{3}^{2}\right)  ^{2}+2a^{2}\left(
x_{0}^{2}+x_{1}^{2}\right)  =0 \label{7}%
\end{equation}
and%
\begin{align}
4x_{0}\left(  x_{0}^{2}+x_{1}^{2}+x_{2}^{2}+x_{3}^{2}\right)  +4a^{2}x_{0}  &
=0,\label{8}\\
4x_{1}\left(  x_{0}^{2}+x_{1}^{2}+x_{2}^{2}+x_{3}^{2}\right)  +4a^{2}x_{1}  &
=0,\label{9}\\
4x_{2}\left(  x_{0}^{2}+x_{1}^{2}+x_{2}^{2}+x_{3}^{2}\right)   &
=0,\label{10}\\
4x_{3}\left(  x_{0}^{2}+x_{1}^{2}+x_{2}^{2}+x_{3}^{2}\right)   &  =0
\label{11}%
\end{align}
for $a\in C,$ $a\neq0$.

To see that there is no non-zero solution of the above system of equations, we
first note that Eq.(\ref{10}) and Eq.(\ref{11}) imply $x_{2}=0$ and $x_{3}=0$
respectively. So either $x_{0}\neq0$ or $x_{1}\neq0$ which, by Eq.(\ref{8})
and Eq.(\ref{9}), imply $a^{2}=-(x_{0}^{2}+x_{1}^{2}).$ But then Eq.(\ref{7})
tells us $-a^{4}=0$ or $a=0$, which contradicts with the sheaf condition that
$a\neq0$. This completes the proof.

Since $\partial_{\nu}\partial_{\mu}h(x_{0},x_{1},x_{2},x_{3})\neq0$ \ for
$x_{\mu}=(0,0,0,0)$, the order of singularity of the sheaf line calculated in
Eq.(\ref{line}) is $2$ and is higher than those of other normal jumping lines.
We note that by using Eq.(\ref{id6}), the jumping line condition is
$\det\Delta^{\circledast}\Delta=\det\beta_{\sigma\lbrack x:y:z:w]}%
\alpha_{\lbrack x:y:z:w]}=0.$ On the other hand, the sheaf line is further
constrained by another condition that $\alpha$ is not injective (or $\beta$ is
not surjective). So it seems to be reasonable to conjecture that in general
the order of singularity of a sheaf line is higher than those of other normal
jumping lines.

\section{Conclusion}

In this paper, we calculate a sheaf line in $CP^{3}$ which is a fiber line on
$S^{4}$ spacetime supporting sheaf points on $CP^{3}$ of Yang-Mills instanton
sheaves for some given ADHM data we obtained previously in \cite{Ann2}. We
found that this sheaf line is a special jumping line over $S^{4}$ spacetime.
Incidentally, we discover a duality symmetry among YM instanton sheaf
solutions with dual ADHM data.

To understand the effect of sheaf line on $S^{4}$ spacetime, we calculate the
singularity structure of the connection $A$ and the field strength $F$ at the
corresponding singular point on $S^{4}$ of this sheaf line. We found that the
order of singularity at the singular point on $S^{4}$ associated with the
sheaf line in $CP^{3}$ is higher than those of other singular points
associated with normal jumping lines. We conjecture that this is a general
feature for sheaf lines among jumping lines.

One unexpected benefit we obtained in our search of the sheaf line was the
great simplification of the calculation of $v$ in Eq.(\ref{ttt}) and the
corresponding connection $A$ and the field strength $F$ in Eq.(\ref{f2})
associated with the sheaf ADHM data. In fact, we have seen that for the sheaf
ADHM data the explicit form of $SL(2,C)$ YM $2$-instanton (or $SU(2)$
\textit{complex} YM $2$-instanton) field strength without removable
singularities can be exactly calculated! To understand the mechanism of this
simplification of the calculation of YM instanton, more explicit examples of
sheaf lines will be helpful.

It will be important to calculate more examples of sheaf lines associated with
YM instanton sheaves for instanton with higher topological charges. However,
it was shown that there has no sheaf line structure for the $SL(2,C)$ diagonal
CFTW $k$-instanton \cite{Ann2}. To explicitly construct YM instanton sheaves,
one needs to first work out explicitly non-diagonal ADHM $k$-instanton
solutions which are in general difficult to calculate for $k>3$.

Surprisingly, in a recent publication \cite{Ann5} an explicit example of
"\textit{extended} complex YM $2$-instanton sheaf" solution with
\textit{diagonal} $y$ ADHM data was found, and the corresponding $2$-instanton
field strength and sheaf line were calculated. Moreover, the results of the
\textquotedblleft sheaf line\textquotedblright\ as a special jumping line (see
Eq.(\ref{ppp})) presented in section III.D and the results of section IV.C.4
which state that the order of singularity for the field strength at the
singular point on $S^{4}$ associated with the sheaf line in $CP^{3}$ is higher
than those of other singular points associated with normal jumping lines, were
explicitly justified by results of this new example. See the details in
Eqs.(4.142) to (4.174) in \cite{Ann5}.

Recently, some examples of YM instanton sheaves with topological charges $3$
and $4$ were discovered and explicitly constructed \cite{Ann3}. The sheaf
lines over $S^{4}$ of these YM instanton sheaves with higher topological
charges and the associated singular structures of $A$ and $F$ are currently
under investigation.

\subsection{Discussion}

It will be important and interesting to find applications of solving $SL(2,C)$
YM instanton (or $SU(2)$ \textit{complex} YM instanton) solutions to real
physical systems. As was mentioned in page $4$ of the Introduction, one
motivation to study $SL(2,C)$ (or in general $SL(n,C)$) SDYM equations is the
central role they played in the integrable systems, such as the sine-Gordon
equation which found numerous applications in many physical systems. For
example, the long Josephson junctions is a device where the sine-Gordon
equation perfectly describes \cite{prb}. In mechanics, the chain of coupled
pendula can also be approximately described by the sine-Gordon equation.

In an integrable system the integrability condition can be realized through
the approach of Lax pairs. For the case of $4D$ SDYM system, one defines%
\begin{equation}
L_{1}=(\partial_{1}+A_{1})+\zeta(\partial_{3}+A_{3}),L_{2}=(\partial_{2}%
+A_{2})+\zeta(\partial_{4}+A_{4})
\end{equation}
where $\zeta$ is the spectral parameter and the YM connection $A_{\mu}$ are
$SL(2,C)$ matrices. It is then easy to see that $[L_{1,}L_{2}]=0$ leads to the
$SL(2,C)$ SDYM equations%
\begin{equation}
F_{12}=F_{34}=F_{14}+F_{32}=0. \label{SDYM}%
\end{equation}
In the above equation Eq.(\ref{SDYM}), the SDYM equations were written in the
null coordinates \cite{cny}%
\begin{equation}
y_{1}=x_{1}+ix_{2}\text{, \ }y_{2}=x_{1}-ix_{2}\text{, \ }y_{3}=x_{3}%
-ix_{4}\text{, \ }y_{4}=x_{3}+ix_{4}.
\end{equation}
\qquad\qquad

On the other hand, as an example, for the case of the sine-Gordon equation,
one defines%
\begin{align}
L_{1}  &  =2\partial_{u}+%
\begin{pmatrix}
f & 0\\
0 & -f
\end{pmatrix}
+\zeta%
\begin{pmatrix}
0 & e^{i\phi/2}\\
e^{-i\phi/2} & 0
\end{pmatrix}
,\\
L_{2}  &  =-2\partial_{v}+\zeta%
\begin{pmatrix}
g & 0\\
0 & -g
\end{pmatrix}
+%
\begin{pmatrix}
0 & e^{-i\phi/2}\\
e^{i\phi/2} & 0
\end{pmatrix}
\end{align}
where $f=f(u,v)$, $g=g(u,v)$ and $\phi=\phi(u,v)$ depending on only two space
parameters. It can be shown that $[L_{1,}L_{2}]=0$ leads to the sine-Gordon
equation%
\begin{equation}
\phi_{uv}+\sin\phi=0.
\end{equation}

Now one well-known result in the integrable systems was that the $2D$
sine-Gordon equation can be obtained by dimensional reduction of $4D$
$SL(2,C)$ SDYM through the above approaches. For a review see the reference
\cite{integr}. It is in this sense, the present authors believe that the
$SL(2,C)$ YM instanton solutions including the instanton sheaf solutions with
the new "integrability" discussed in the paragraphes below Eq.(\ref{fff}) may
find interesting applications in the real physical systems. More works need to
be done to uncover these connections.

Finally, we would like to point out that we choose to work on $R^{4}$ in which
our complex instantons live. More precisely our complex instantons were
studied on $S^{4}$ and by conformal transformations these solutions were
converted on $R^{4}$. The space $CP^{3}$, despite that it is a curved space,
serves only as an auxiliary space which is put to use for purely mathematical
reasons, namely, for the Penrose-Ward transform--our main tool which together
with the ADHM method helps to solve the SDYM equation--to work well. One may
be wondering whether this methodology can be adopted in other topological
spaces. Indeed, the idea of Penrose-Ward transform has been suitably
generalized to other spaces \cite{42}. See Theorem 5.2 in p. 441 for further
information. A concrete case using ADHM techniques on the hyperbolic space has
been studied in a recent publication \cite{popov}; see also \cite{kahler}.

\begin{acknowledgments}
The work of J.C. Lee is supported in part by the Ministry of Science and
Technology and S.T. Yau center of NCTU, Taiwan. The work of I-H. Tsai is
supported by the Ministry of Science and Technology of Taiwan under grant
number 105-3114-C-002-001-ES. We thank the anonymous referee who suggested
addressing the issues in the subsection "Discussion" of section V. Conclusion.
\end{acknowledgments}

\appendix%

%TCIMACRO{\TeXButton{equation number in appendix}{\setcounter{equation}{0}
%\renewcommand{\theequation}{\thesection.\arabic{equation}}}}%
%BeginExpansion
\setcounter{equation}{0}
\renewcommand{\theequation}{\thesection.\arabic{equation}}%
%EndExpansion


\begin{thebibliography}{99}
\bibitem {misleading}This subsection is the corrected version of the Appendix
in Ref [12].
\end{thebibliography}
\end{document}